\documentstyle[12pt,epsfig]{article}
\normalbaselines
\catcode `\@=11
\@addtoreset{equation}{section}

\def\section{\@startsection {section}{1}{\z@}{-3.5ex plus -1ex minus
     -.2ex}{2.3ex plus .2ex}{\normalsize\bf}}
\def\subsection{\@startsection{subsection}{2}{\z@}{-3.25ex plus -1ex minus
 -.2ex}{1.5ex plus .2ex}{\normalsize\sl}}

\def\thebibliography#1{\section*{References\markboth
  {REFERENCES}{REFERENCES}}\list
  {[\arabic{enumi}]}{\settowidth\labelwidth{[#1]}\leftmargin\labelwidth
  \advance\leftmargin\labelsep
  \usecounter{enumi}}
  \def\newblock{\hskip .11em plus .33em minus -.07em}
   \sloppy
  \sfcode`\.=1000\relax}
 
\catcode `\@=12

\skip\footins = 14pt 
\begin{document}
\noindent
\
\hfil June 2000 
\break
-----------------------------------------------------------------------------------------------------------------

\vspace*{2.2cm}
\noindent
{\bf FERMION QUANTUM NUMBERS AND FAMILIES REPLICATION} \\[1mm]  
\noindent
{\bf FROM AN EXTENSION OF SPACE-TIME RELATIVITY}\vspace{1.3cm}\\
\noindent
\hspace*{1in}
\begin{minipage}{13cm}
P. Maraner \vspace{0.3cm}\\
\makebox[3mm]Dipartimento di Fisica, Universit\`a di Parma\\
\makebox[3mm]Parco Area delle Scienze, I-43100 Parma, Italy
\end{minipage}

\vspace*{0.5cm}
\begin{abstract}
\noindent
The fermionic sector of the Standard Model of Elementary Particles 
emerges as the low energy limit of a single fermionic field freely 
propagating in a higher dimensional background. 
The local geometrical framework is obtained by enforcing at a space-time 
level the whole gauge group $SO(1,3)\times U(1)\times SU(2)\times SU(3)$ 
associated to fundamental interactions; equivalently, by assuming that 
internal gauge transformations are indeed local space-time transformations. 
The geometry naturally embodies freedoms corresponding to gravitational 
and non-gravitational gauge fields. As a consequence of the fact 
that the structural group is in part unitary, the motion of test particles
gets automatically squeezed on an effective 1+3 space-time. 
Dimensional reduction takes place without compactification.
In close analogy to the special relativistic mass-energy relation, the theory 
associates to every elementary particle an intrinsic energy presumably
of the order
of the Planck scale. The theory predicts the existence of a right-handed
component $\nu_R$ of the neutrino and indicates the possibility of an extra 
$U(1)$ gauge interaction.
\end{abstract}

{\sl PACS:} 12.10.-g, 04.50.+h
\vskip0.5cm
\noindent
The General Theory of Relativity describes gravitational phenomena 
in terms of two basic concepts: {\it space-time} and {\it matter}.
Matter propagates freely in space-time while the structure of space-time  
is determined by the matter distribution.
This picture represents the maximal reduction process ever achieved in 
a physical theory, freeing us at once of the prejudice of an absolute 
rigid space-time and of the effective concept of a gravitational field 
defined {\it on} space-time. 
All gravitational phenomena find a realistic comprehension in terms
of operative geometrical concepts.
 There is a single respect under which the theory appears incomplete.
While space-time is entirely characterized by the
framework of $SO(1,3)$ pseudo-Riemannian geometry we do not have an 
elementary characterization of matter.
Given the existence of non-gravitational phenomena, this 
is perfectly reasonable if not even desirable.
In any case, the general picture is so attractive from a philosophical 
viewpoint that we crave an analogue comprehension of all fundamental 
interactions. 
In this paper we argue that this is indeed possible --at least at a classical 
level-- at the price of a further extension of space-time relativity. 
 The resulting theory --possibly describing gravitational, 
electromagnetic, weak and strong forces as effects of the local geometrical 
structure of an extended space-time-- displays a certain degree of 
completeness. A single 
fermionic matter field freely propagating in a background gets automatically 
squeezed on a $1+3$ effective space-time realizing 
all basic feature of elementary matter fields as described by the Standard 
Model of Elementary Particle before electro-weak symmetry breaking 
({\it i.e.}~the existence of chiral fermions, the relative hypercharge, 
isospin and color quantum numbers, the phenomenon of families replication as 
well as elementary matter field equations). 

\subsection*{Gauge transformations as local space-time transformations}
The geometrical framework of General Relativity grounds on two basic
assumption inferred from experience:
Lorentz Invariance, the physical equivalence of all space-time reference 
frames related by Lorentz transformations, and the 
Principle of Equivalence, asserting the possibility of eliminating the 
effects of gravitation in an infinitesimal neighborhood of every point 
by an appropriate choice of coordinates.
 The former fixes the rigid properties of space-time in the absence of 
gravitational phenomena; it determines the local gauge group $SO(1,3)$ or, 
equivalently, the Minkowskian metric $\eta_{\alpha\beta}$. 
 The latter identifies gravitational phenomena with a non trivial parallel 
transport of space-time reference frames forcing us to relax the rigidity 
hypothesis and fixing the {\it local} space-time structure in terms of 
$SO(1,3)$ pseudo-Riemannian geometry. 
Technically one proceeds by promoting the Minkowskian metric   to a point 
dependent metric $\eta_{\alpha\beta}\rightarrow g_{\mu\nu}(x)$ satisfying 
the maximal compatibility condition $\nabla_\kappa g_{\mu\nu}=0$.  
Lorentz Invariance and the Principle of Equivalence allows us to understand
gravitational phenomena in pure terms of space-time geometry.
 How to generalize these basic assumptions to take into account 
non-gravitational interactions? 
 If we insist with a geometrical picture we should focus on geometrical
properties. 
 It is well known that the only geometrical property shared by all fundamental 
interactions is gauge invariance. This basic fact is at the heart of 
most attempts of breaking through toward a unified picture. 
However, the emphasis always has been on gauge invariance alone or 
geometry alone rather than on gauge invariance as a local geometric 
property of space-time.
There is a deep asymmetry in nowadays comprehension of gauge invariance 
associated to gravitational and non-gravitational interactions. 
On one side we have a clear understanding of the local $SO(1,3)$ 
gravitational gauge invariance in terms of observer's freedom of choosing 
at will an orthonormal reference frame in every space-time point. 
To the side of non-gravitational interactions, gauge invariance enters in a 
rather technical way, 
guarantying renormalizability and hence perturbative consistence of the 
quantum  theory. Nonetheless, the physical meaning of the `internal' 
$U(1)\times SU(2)\times SU(3)$ gauge invariance associated 
to electromagnetic, weak and strong forces remains obscure.
 Attempts of shedding light on this point date back to the work of
Weyl, Kaluza and Klein in the twenties reaching present day speculations on
supergravity and strings. It is curious to note that even though most of
these theories introduce extra dimensions besides the four familiar ones,
no serious attempts have been made to assimilate `internal' gauge
invariance with the local freedom associated to the choice of an
orthonormal reference frame in extra dimensions --simply in line with
gravity. While gravitational gauge invariance is a robust (insensitive
to continuous deformations) local property of space-time,
`internal' gauge invariance is in general assimilated with fragile
global properties of particular highly symmetrical solutions. 
 We do find such a deep asymmetry  in the comprehension of the
only geometrical property surviving in our effective description of
fundamental interactions  rather unattractive. 
If we do accept as a working hypothesis the introduction of extra dimensions 
we should keep in mind that we are already adding to the theory an extra 
gauge freedom completely analogous to the gravitational gauge freedom. 
Why not identifying this freedom with non-gravitational gauge freedom? \\
In support of this elementary geometrical argument there is a well known
physical fact. Besides the generic consideration that all fundamental 
interactions share gauge invariance, it is a very remarkable 
experimental fact that gauge groups realized in nature are either 
pseudo-orthogonal or unitary.
Orthogonal and unitary groups are intimately connected, playing the identical
role of tangent space structural groups in real and complex geometry 
respectively. This fact indicates us clearly the way to take. 
The `internal' gauge group $U(1)\times SU(2)\times SU(3)$ should 
simply appear on the side of the gravitational gauge group $SO(1,3)$ as 
--possibly part of-- a larger space-time structural group\footnote{The 
possibility of constructing differential geometries with generic structural 
groups was considered by S.~Weinberg in a Kaluza-Klein context 
\cite{Weinberg}. The geometry we consider in this paper is 
a particular case of Weinberg's generalization. However, it will be excessive 
to consider our geometrical structure as a generalized differential
geometry; it is rather an ``interpolation'' between real and 
complex classical differential geometries.}. 
In different words, we surrender to the experimental evidence that the local 
gauge group associated to fundamental interactions is $SO(1,3)\times U(1)\times
SU(2)\times SU(3)$ and we enforce this at a space-time level by extending
the request of space-time Lorentz Invariance to the one of 
space-time {\sl General Gauge Invariance}.
 This amounts to introduce ten extra space-time dimensions of a somehow 
`complex' nature besides  the four familiar real ones. In the absence 
of interactions the real and complex nature of different space-time directions
is characterized by an appropriate rigid metric structure generalizing
the Minkowskian structure.
 By assuming the validity of a {\sl General Principle of Equivalence} 
for all fundamental interactions we are then led to relax the rigidity
conditions identifying all fundamental interactions with a non-trivial 
parallel transport of space-time reference frames. Technically this is 
obtained again by promoting the rigid space-time metric to a point dependent 
metric satisfying  a maximal compatibility condition.
General Gauge Invariance and the General Principle of Equivalence
bring us to a unified description of fundamental interactions.
This is not only in the mere technical sense that we succeed in recasting 
all fundamental force fields in a single geometrical object but on the deeper 
physical and philosophical ground that all interactions appear as identical 
operative consequences of a non trivial parallel transport of space-time 
reference frames. 
 In the first part of this paper we reconstruct the local geometrical 
structure of space-time out of these two basic assumption.

\subsection*{On the four dimensional nature of space-time and extra dimensions}
 The generalization of a fundamental concept like  the one of space-time 
is clearly an extremely delicate task that we have the right to perform 
only if bringing a concrete improvement in our understanding of Nature.
 Therefore, in introducing extra space-time dimensions and a particular 
hypothesis on their nature, we have to immediately face two urgent questions. 
 First of all, we do have to explain why our direct perception of space-time 
is limited to four dimensions.
 Secondly, but of a major importance, we do have to display an unmistakable
trace left by extra dimensions in the effective four dimensional physics.
 The traditional approach to the first point is that of introducing
additional hypothesis  on the topological nature of extra dimensions
--hence on the global structure of space-time-- assuming them to
dynamically compactify on extremely small length scales. 
Basic experimental evidence on chiral fermions is apparently against 
this hypothesis in pure metrical theories \cite{Wetterich-Witten}. In any case 
there is no answer to the second point in the traditional approach.
 It is instead an intrinsic property of the local geometrical structure 
we have been lead to
--given essentially by the fact that the complex part of the space-time 
connection couples to matter like a magnetic field--
that the free motion of matter in a background gets automatically squeezed 
on a real $1+3$ effective space-time.
This gives a natural answer to the question of dimensional reduction without 
the introduction of any subsidiary hypothesis. In addition the trace left by 
extra dimensions in the effective four dimensional physics appears 
unmistakably in the structure of elementary matter fields. 

\subsection*{The fermionic sector of the Standard Model of Elementary 
Particles}
The structure of elementary matter fields before electro-weak symmetry 
breaking is a basic building block of the Standard Model of Elementary 
Particles. As extracted from experience the fundamental fermionic 
representation  appears complicated and redundant. 
Complicated because the basic pattern --a family-- consist of five irreducible 
representations of $U(1)\times SU(2)\times SU(3)$ carrying a definite 
chirality: 
a right-handed 
isospin singlet ${\sf I}={\mathbf 1}$,
color singlet   ${\sf C}={\mathbf 1}$,
of hypercharge ${\sf Y}=1$
identified with the right-handed electron $e_R$;
a left-handed 
isospin doublet ${\sf I}={\mathbf 2}$,
color singlet   ${\sf C}={\mathbf 1}$,
of hypercharge ${\sf Y}=1/2$
which components correspond to the left-handed electronic neutrino $\nu_{eL}$
and electron $e_L$;
two right-handed 
isospin singlet ${\sf I}={\mathbf 1}$,
color triplets  ${\sf C}={\mathbf 3}$,
of hypercharge ${\sf Y}=1/3$ and ${\sf Y}=-2/3$
identified with the right-handed quarks down $d_R$ and up $u_R$ respectively;
and a left-handed 
isospin doublet ${\sf I}={\mathbf 2}$,
color triplets  ${\sf C}={\mathbf 3}$,
of hypercharge ${\sf Y}=-1/6$ corresponding to the left-handed components of 
the quarks down $d_L$ and up $u_L$. In a more synthetic notation we can
write these fifteen elementary matter fields as
\[
\begin{array}{llll}
\ \ e_R=({\mathbf 1},{\mathbf 1})_{+1}^+ & 
\left(\begin{array}{r} \nu_{eL}\cr e_L \end{array}\right)=
                   ({\mathbf 2},{\mathbf 1})_{+1/2}^- &
\begin{array}{l}
u_R=({\mathbf 1},{\mathbf 3})_{-2/3}^+ \\[2mm]
d_R=({\mathbf 1},{\mathbf 3})_{+1/3}^+ 
\end{array} &
\left(\begin{array}{r}  u_L\cr d_L \end{array}\right)=
                    ({\mathbf 2},{\mathbf 3})_{-1/6}^-
\end{array}
\]
The fundamental representation is redundant because this basic pattern is 
repeated --at least-- three times in the generations corresponding to the 
leptons $\mu$, $\nu_{\mu}$ and the quarks charm $c$ and strange $s$
\[
\begin{array}{llll}
\ \ \mu_R=({\mathbf 1},{\mathbf 1})_{+1}^+ & 
\left(\begin{array}{r} \nu_{\mu L}\cr \mu_L \end{array}\right)=
                   ({\mathbf 2},{\mathbf 1})_{+1/2}^- &
\begin{array}{l}
c_R=({\mathbf 1},{\mathbf 3})_{-2/3}^+ \\[2mm]
s_R=({\mathbf 1},{\mathbf 3})_{+1/3}^+ 
\end{array} &
\left(\begin{array}{r}  c_L\cr s_L \end{array}\right)=
                    ({\mathbf 2},{\mathbf 3})_{-1/6}^-
\end{array}
\]
and to the leptons $\tau$, $\nu_\tau$ and the quarks top $t$ and bottom $b$
\[
\begin{array}{llll}
\ \ \tau_R=({\mathbf 1},{\mathbf 1})_{+1}^+ & 
\left(\begin{array}{r} \nu_{\tau L}\cr \tau_L \end{array}\right)=
                   ({\mathbf 2},{\mathbf 1})_{+1/2}^- &
\begin{array}{l}
t_R=({\mathbf 1},{\mathbf 3})_{-2/3}^+ \\[2mm]
b_R=({\mathbf 1},{\mathbf 3})_{+1/3}^+ 
\end{array} &
\left(\begin{array}{r}  t_L\cr b_L \end{array}\right)=
                    ({\mathbf 2},{\mathbf 3})_{-1/6}^-
\end{array}
\]
As a further assumption the Standard Model of Elementary Particles
prescribes elementary matter fields to minimally interact with
the gauge potentials ${\sf A}_\mu^y$, ${\sf A}_\mu^{\iota{\sf i}}$, ${\sf i}=
1,2,3$, and ${\sf A}_\mu^{\kappa{\sf c}}$, ${\sf c}=1,2,...,8$, describing
hyper $U(1)$, weak $SU(2)$ and strong $SU(3)$ interactions. In a standard
notation matter field equations read
\[  
\gamma^\mu(-i\partial_\mu
-{\sf Y}_{\sf p}{\sf A}^y_\mu
-{\sf A}^{\iota{\sf i}}{\sf I}^{\sf i}_{\sf p}
-{\sf A}^{\kappa{\sf c}}{\sf C}^{\sf c}_{\sf p})
           \psi_{\sf p} =0
\]
where $\psi_{\sf p}$ is any of the particle listed above, 
${\sf Y}_{\sf p}$ the relative hypercharge and ${\sf I}^{\sf i}_{\sf p}$, 
${\sf C}^{\sf c}_{\sf p}$ the relative isospin and color representations
(multiplet and spin indices are understood). \\
Some light on the single family structure is shed by the observation
\cite{Georgi-Glashow} 
that --while $U(1)\times SU(2)\times SU(3)$ naturally embeds in $SU(5)$
and this is $SO(10)$-- the left-handed fermionic representation 
$({\mathbf 1},{\mathbf 1})_{-1}^- \oplus
({\mathbf 2},{\mathbf 1})_{+1/2}^- \oplus
({\mathbf 1},\bar{\mathbf 3})_{+2/3}^- \oplus
({\mathbf 1},\bar{\mathbf 3})_{-1/3}^-\oplus
({\mathbf 2},{\mathbf 3})_{-1/6}^-$
matches the simpler $SU(5)$ representation
$\bar{\mathbf 5}^-+{\mathbf 10}^-$; if we then accept to include as 
a further fundamental fermion a highly non-interactive right-handed
neutrino $\nu_R=({\mathbf 1},{\mathbf 1})_{\ 0}^+$ the total left-handed 
fermionic representation recasts in to the single $SO(10)$
representation ${\mathbf 16}^-$.  Standard gauge theories based on
$SU(5)$ or $SO(10)$ --not including gravity-- pay the simplification 
of gauge group and  fundamental fermionic representation at the very 
high price of introducing several unobserved vector mesons other than  
a whole army of Higgs bosons necessary to break down
the symmetry to the observed one. In addition they leave unsolved 
the problem of family replication and give no contribution to our
understanding of `internal' gauge invariance. From our viewpoint,
we observe that the natural embedding of gauge group
and matter representation in $SO(10)$ is perfectly compatible
with --and perhaps a further indication of-- the introduction of 
ten extra dimensions. \\
In the second and third part of this paper we analyze the free
motion of a fermion in a given background. The fermionic sector
of the Standard Model of Elementary Particles emerges as the low
energy limit of the theory.

\newpage
\section{Space-time structure}

The geometrical structure of General Relativity is that of $SO(1,3)$
pseudo-Riemannian geometry. Space-time is described 
by a {\it real manifold}. Gravitational forces  by a  {\it metric 
connection} with tangent space structure
\[
\eta_{\alpha\beta}=
\left(
\begin{array}{cccc}
-1&   &   &   \\
  & 1 &   &   \\
  &   & 1 &   \\
  &   &   &  1
\end{array}
\right)
\]
The gravitational field can be identified with different geometrical 
objects: the metric tensor $g_{\mu\nu}$, the affine connection 
$\Gamma_{\mu\nu}^\rho$ or the spin connection $\Omega_{\kappa,\alpha\beta}$.
The vanishing of space-time torsion --as required by the Principle of 
Equivalence-- makes the three structure to carry the same amount of
geometrical information. 
 The gauge character of gravitational interactions is most easily displayed 
in terms of spin connection.
 A fundamental notion in the theory is that of {\it reference frame}. 
A reference frame at a point $x$ is a set of three orthonormal space-like 
vectors $\left.{e_1^{\ \mu}}\right|_x, \left.{e_2^{\ \mu}}\right|_x,
\left.{e_3^{\ \mu}}\right|_x$ plus a clock. Geometrically the clock is 
interpreted as a time-like unit vector $\left.{e_0^{\ \mu}}\right|_x$ which 
is conveniently chosen orthogonal to the first three.
An assignation of a reference frame in every space-time point is therefore 
a set of four vector fields satisfying
\[ 
e_\alpha^{\ \mu}e_\beta^{\ \nu}g_{\mu\nu}=\eta_{\alpha\beta}
\] 
We see that a reference frame is determined up to a point dependent 
Lorentz pseudo-rotation, 
\[
e_\alpha^{\ \mu}\rightarrow \Lambda_\alpha^{\ \beta}(x)e_\beta^{\ \mu}
\] 
where in every point $\Lambda_\alpha^{\ \gamma}\Lambda_\beta^{\ \delta}
\eta_{\gamma\delta}=\eta_{\alpha\beta}$.  
In order to compare physical phenomena in different space-time points we
are lead to construct the (spin) connection
\[
 \Omega_{\kappa,\alpha\beta}= 
 e_\alpha^{\ \mu}\left(\nabla_\kappa e_\beta^{\ \nu}\right)g_{\mu\nu}
\]
where $\nabla$ denotes the covariant derivative induced by the metric 
$g_{\mu\nu}$ (on curved indices only). The  connection is clearly
anti-symmetric in the indices $\alpha$ and  $\beta$,  belonging therefore 
to the Lie algebra $so(1,3)$.
Under a local $SO(1,3)$ transformation
of reference frames
\[
\Omega_{\kappa,\alpha\beta} \rightarrow
\Lambda_\alpha^{\ \gamma}\Lambda_\beta^{\ \delta}\Omega_{\kappa,\gamma\delta}
+\Lambda_\alpha^{\ \gamma}
 \left(\partial_\kappa \Lambda_\beta^{\ \delta}\right)\eta_{\gamma\delta}
\]
which clearly display the gauge nature of the gravitational field.
The freedom of choosing at will a reference frame in every  space-time 
point is the deep physical meaning of the $SO(1,3)$ gauge invariance 
of gravitational forces.

\vskip0.2cm
\noindent
We are willing to enforce the whole gauge group $SO(1,3)\times U(1)\times 
SU(2)\times SU(3)$ associated to fundamental interactions at a space-time
level. This requires the introduction of ten complex dimensions
besides the four familiar real ones. Therefore, it is convenient to start by
briefly recalling the main feature of complex geometry.

\vskip0.2cm
\noindent
There are essentially two equivalent approaches to the definition of 
a {\it complex manifold}. The first one is repeating the definition of real 
manifold in terms of local coordinates and transformation functions,
with real vector spaces replaced by complex ones and real smooth
functions by homomorphic functions. The second one --that better adapts 
to our task-- is starting from a real manifold and introducing a 
complex structure on it. 
 A complex structure $I_i^{\ j}$ is a tensor field acting on the 
tangent space at every point of the manifold as the multiplication by 
the imaginary unit $i$. More explicitly $I_i^{\ k}I_k^{\ j}=-\delta_i^j$ 
plus an integrability condition necessary to guarantee that imaginary units 
defined in different points fit together consistently. 
 A {\it complex metric connection} on a complex manifold is introduced through
a Riemannian metric $g_{ij}$ preserving the complex structure,
$g_{ij}=I_i^{\ k} I_j^{\ l} g_{kl}$. To  $g_{ij}$ is then 
associated an antisymmetric two-form $\omega_{ij}=I_i^{\ k}g_{kj}$.
Metric and antisymmetric two-form completely characterize the geometry
and are  simply recasted in real and imaginary
part of a single Hermitian metric $h_{ij}=g_{ij}+i\omega_{ij}$.
 Observe that to such a geometry is in general associated a 
non-vanishing torsion proportional to the exterior derivative
of $\omega_{ij}$. Further requiring the vanishing of torsion
makes metric and skew-symmetric two-form maximally compatible,
$\nabla_k\omega_{ij}\equiv0$. Such a geometry plays the role of  
Riemannian geometry in the complex realm and is known as K\"ahlerian
geometry. On the tangent space at every point we always can choose 
coordinates in such a way that $g_{ij}$ and $\omega_{ij}$
take the standard form 
\[
\delta_{ab}=
\left(
\begin{array}{ccc}
1 & 0 &     \\
0 & 1 &     \\
  &   & \ddots
\end{array}
\right)
\ \ \ \ \ \ \ \ \ \ \ 
\varepsilon_{ab}=
\left(
\begin{array}{ccc}
0 & 1  &     \\
-1& 0 &     \\
  &   &   \ddots
\end{array}
\right)
\]
where dots indicate a certain number of blocks of the same type. 
These matrices specify the tangent space structure of the geometry,
playing the very same role of the Minkowskian metric in 
pseudo-Riemannian geometry.
We are now ready to go back to the notion of reference frame. An
assignation of a reference frame in every point of a K\"ahlerian
manifold is a maximal set of independent vector fields $e_a^{\ i}$ 
fulfilling the conditions
\[
e_a^{\ i}e_b^{\ j}g_{ij}=\delta_{ab}
\ \ \ \ \ \ \ \ \ \ \ 
e_a^{\ i}e_b^{\ j}\omega_{ij}=\varepsilon_{ab}
\]
Again, we see that a reference frame is determined up to transformations
\[
e_a^{\ i}\rightarrow U_a^{\ b}(x)e_b^{\ i}
\]
that satisfies in every point $U_a^{\ c}U_b^{\ d}\delta_{cd}=\delta_{ab}$ and  
$U_a^{\ c}U_b^{\ d}\varepsilon_{cd}=\varepsilon_{ab}$. The first condition 
ensures $U_a^{\ b}$ to be  an orthogonal transformation, the second a 
symplectic one. The intersection of orthogonal and symplectic groups 
is the unitary group $U(n)=SO(2n)\cap Sp(2n)$. On a K\"ahlerian manifold
a reference frame is therefore determined up to an unitary transformation.
Comparing informations in different space-time points involves again
the  connection
\[
 \Omega_{k,ab}= 
 e_a^{\ i}\left(\nabla_k e_b^{\ j}\right)g_{ij}
\]
The requirement of a vanishing torsion makes $\Omega_{k,ab}$ to
belong to the Lie algebra $u(n)$. Under a local unitary
transformation $\Omega_{k,ab}$ transform as a $U(n)$ gauge
connection
\[
\Omega_{k,ab}\rightarrow 
U_a^{\ c}U_b^{\ d}\Omega_{k,cd}+
U_a^{\ c}\left(\partial_kU_b^{\ d}\right)\delta_{cd}  
\]
However, $\Omega_{k,ab}$ can not be associated to 
a non-gravitational gauge field. The indices $k$, $a$ and $b$ have
the same transformation properties. Somehow we need a structure
supporting objects having both real space-time and complex gauge indices. 

\vskip0.2cm
 We have all the necessary ingredients to build up such a generalization. 
In analogy with real Riemannian and complex K\"ahlerian geometries we start 
with a space-time described by a real manifold. 
Enforcing General Gauge Invariance we introduce a tangent space structure 
having as isometries the gauge group of fundamental interactions. 
In accordance with the General Principle of Equivalence 
we then construct the local geometrical framework based on this structure. 
Fundamental interactions will be identified with appropriate components of 
the total connection.

\subsection*{General Gauge Invariance and tangent space structure}

We may assume a generic tangent space structure to be given by a 
Hermitian matrix $\hat\eta_{AB}+i\hat\varepsilon_{AB}$. 
 The real symmetric part $\hat\eta_{AB}$ always can be chosen diagonal with 
all  eigenvalues $\pm1$.
 The imaginary anti-symmetric part $\hat\varepsilon_{AB}$ always can be chosen
two by two block diagonal. For its eigenvalues we have many different options.
 When all eigenvalues equal zero we obtain a (pseudo-)Riemannian real 
structure. 
 When all eigenvalues equal $\pm i$ we obtain a (pseudo-)K\"ahlerian complex 
structure.
In the general case we have a group or real (pseudo-)Riemannian
directions corresponding to null eigenvalues, plus one or more groups 
of equivalent complex (pseudo-)K\"ahlerian directions corresponding to 
possibly degenerate complex eigenvalues. 
 The case that better fits to the group $SO(1,3)\times U(1)\times SU(2)\times 
SU(3)$ is obviously the latter. \\
 General Gauge Invariance brings us to infer that space-time has
$4+10$ dimensions and  --in the absence of fundamental interactions--
a rigid metric structure given by $\hat{\eta}_{AB}+i\hat{\varepsilon}_{AB}$, 
where the real part $\hat{\eta}_{AB}$ is the Minkowskian metric
\[
\hat{\eta}_{AB}=
\mbox{diag}(-c^2,\ 1,\ 1,\ 1,\ 1,\ 1,\ 1,\ 1,\ 1,\ 1,\ 1,\ 1,\ 1,\ 1)
\hskip2.4cm
\]
and the imaginary part $\hat{\varepsilon}_{AB}$ is the antisymmetric two-form
\[
\hat{\varepsilon}_{AB}=
\left(
\begin{array}{cccrcccclccccc}
0 & 0 &   &     \vline&   &   &   &   &              &   &   &   &    &   \\ 
0 & 0 &   &     \vline&   &   &   &   &              &   &   &   &    &   \\ 
  &   & 0 &0\ \ \vline&   &   &   &   &              &   &   &   &    &   \\ 
  &   & 0 &0\ \ \vline&   &   &   &   &              &   &   &   &    &   \\ 
\cline{1-8}
  &   &   & \vline    & 0 & w &   &   &\vline        &   &   &   &    &   \\ 
  &   &   & \vline    &-w & 0 &   &   &\vline        &   &   &   &    &   \\
  &   &   & \vline    &   &   & 0 & w &\vline        &   &   &   &    &   \\
  &   &   & \vline    &   &   &-w & 0 &\vline        &   &   &   &    &   \\ 
\cline{5-14}
  &   &   &           &   &   &   &   &\vline\ \ \ 0 & s &   &   &    &   \\
  &   &   &           &   &   &   &   &\vline\  -s   & 0 &   &   &    &   \\ 
  &   &   &           &   &   &   &   &\vline        &    & 0 & s &   &   \\
  &   &   &           &   &   &   &   &\vline        &    &-s & 0 &   &   \\
  &   &   &           &   &   &   &   &\vline        &    &   &   & 0 & s \\
  &   &   &           &   &   &   &   &\vline        &    &   &   &-s & 0
\end{array}
\right)
\]
$c$ is at usual the speed of light and $w$, $s$ two real positive parameters 
not fixed by present considerations\footnote{A word about notation. 
We have quite a large number of objects and different indices. 
We refer as {\it ordinary space-time directions} the four real directions.
All other directions are collectively addressed  as {\it extra space-time 
directions}.
The ones corresponding to eigenvalues $\pm iw$ are called  
{\it weak directions} while the ones corresponding to $\pm is$  
{\it strong directions}.
As a rule we put a hat $\hat{\ }$ over all  $14$-dimensional objects. 
This allows us to use the same symbol to denote the same geometrical object
in different contexts. 
Capital Latin letters form $A$ run from $0$ to $13$ and 
denote space-time flat-indices. 
Greek letters  from $\alpha$ denote ordinary space-time 
flat-indices and take values $0$, $1$, $2$, $3$. 
Latin letters from $a$ take values $4, ..., 13$ and denote 
flat-indices in weak and strong directions. We also split
this set of indices in $a_w$ going from $4$ to $7$ and $a_s$ going
from $8$ to $13$. Furthermore, we find it convenient to introduce the
notation $1_w=4$, $2_w=5$, $3_w=6$, $4_w=7$ and $1_s=8$, ..., $6_s=13$.
The same type of letters  from the mid of the alphabet
will denote the corresponding curved-indices.}. 
Observe that while real directions may 
be rescaled by setting --as we do-- $c=1$, the same is not possible for single 
complex directions. As an example, rescaling the fourth and fifth coordinates 
(we count from zero) in such a way that $\hat\varepsilon_{45}=
-\hat\varepsilon_{54}=1$ will introduce $w$ in $\hat\eta_{44}$ and 
$\hat\eta_{55}$. In the following we  assume $w\neq s$,
\footnote{We may equally be tempted to assume all the five complex 
directions as equivalent, obtaining an attractive grand unified 
$U(1)\times SU(5)$ gauge group. However, we find more economic to 
break space-time symmetry --it is already broken-- at a fundamental level by 
assuming the existence of different groups of inequivalent complex directions.
 At the end of the story this will avoid the introduction a large number of 
unobserved vector mesons and corresponding Higgs scalars necessary 
to break the group down to the desired symmetry.}
that is, two groups of inequivalent complex directions. 

The role of Lorentz transformations is now taken by the isometries
of the local Hermitian structure $\hat\eta_{AB}+i\hat\varepsilon_{AB}$. 
In different words we claim the physical equivalence of all space-time 
reference frames related by transformations $\hat\Lambda_A^{\ B}$ fulfilling 
conditions
\[ 
\hat\Lambda_A^{\ C}\hat\Lambda_B^{\ D}\hat\eta_{CD}=\hat\eta_{AB}
\ \ \ \mbox{and}\ \ \ 
\hat\Lambda_A^{\ C}\hat\Lambda_B^{\ D}\hat\varepsilon_{CD}=
                                  \hat\varepsilon_{AB}  
\]
By dividing $\hat\Lambda_A^{\ B}$ in appropriate blocks we realize 
that all non diagonal blocks have to equal zero 
\[
\hat\Lambda_A^{\ B}=
\left(
\begin{array}{rcl}
\Lambda_\alpha^{\ \beta}\ \ \vline &              &        \\[1mm]
\cline{1-2}
\vline& &\\[-4mm]
                       \vline &\!\! W_{a_w}^{\ {b_w}} \!\!&\vline  \\[1mm]
\cline{2-3}
&&\vline\\[-4mm]
                         &                        &\vline\ \ S_{a_s}^{\ {b_s}}
\end{array}
\right)
\]
while $\Lambda_\alpha^{\ \beta}\in SO(1,3)$, $W_{a_w}^{\ {b_w}}\in U(2)$ 
and  $S_{a_s}^{\ {b_s}}\in U(3)$. The isometry group is therefore the direct 
product $SO(1,3)\times U(1)\times U(1)\times SU(2)\times SU(3)$. 
The extension includes  naturally the pattern of local symmetries of 
fundamental interactions. The only {\sl extracharge} we have to pay is an 
additional $U(1)$ gauge symmetry. \\
 For later use and also to fix notation it convenient to explicitly
write down the conditions fulfilled by  infinitesimal transformations. 
Rewriting $\hat\Lambda_A^{\ B}=\delta_A^B +\hat\lambda_{A}^{\ B}+...$,
with $\hat\lambda_{A}^{\ B}$ taken infinitesimal
and using the conditions above we obtain
\[
\hat\lambda_{A}^{\ C}\hat\eta_{CB}+
\hat\lambda_{B}^{\ C}\hat\eta_{CA}=0
\ \ \ \mbox{and}\ \ \ 
\hat\lambda_{A}^{\ C}\hat\varepsilon_{CB}-
\hat\lambda_{B}^{\ C}\hat\varepsilon_{CA}=0
\]
Dividing ordinary space-time, weak and strong  indices we find that:
\[
\hat\lambda_{\alpha\beta}=-\hat\lambda_{\beta\alpha}
\hskip8cm
\] 
are the usual infinitesimal generators of ordinary 
{\sl Lorentz transformations}; 
\[
\hat\lambda^x=
2w(\hat\lambda_{1_w2_w}+\hat\lambda_{3_w4_w})+ 
2s(\hat\lambda_{1_s2_s}+
   \hat\lambda_{3_s4_s}+\hat\lambda_{5_s6_s})
\hskip1.15cm
\]
and
\[
\hat\lambda^y=
2w(\hat\lambda_{1_w2_w}+\hat\lambda_{3_w4_w})-
2s(\hat\lambda_{1_s2_s}+
   \hat\lambda_{3_s4_s}+\hat\lambda_{5_s6_s})
\hskip1.15cm
\] 
generate respectively a $U(1)$ rotation of all complex directions 
and a relative $U(1)$ rotation of the two blocks of inequivalent complex
directions. These generators of space-time transformations will later be 
identified with the additional {\sl extracharge} and with {\sl hypercharge};
\[
\begin{array}{l}
\hat\lambda^{\iota_1}={1\over2}
 \left(\hat\lambda_{1_w4_w}-\hat\lambda_{2_w3_w}\right) \\[2mm]
\hat\lambda^{\iota_2}={1\over2}
 \left(\hat\lambda_{1_w3_w}+\hat\lambda_{2_w4_w}\right) \\[2mm]
\hat\lambda^{\iota_3}={1\over2}
 \left(\hat\lambda_{1_w2_w}-\hat\lambda_{3_w4_w}\right)
\end{array}
\hskip5.9cm
\]
generates space-time $SU(2)$ transformations that will appear in the 
effective four dimensional physics as {\sl isospin transformations};
\[
\begin{array}{l}
\hat\lambda^{\kappa_1}={1\over2}
 \left(\hat\lambda_{1_s4_s}-\hat\lambda_{2_s3_s}\right) \\[2mm]
\hat\lambda^{\kappa_2}={1\over2}
 \left(\hat\lambda_{1_s3_s}+\hat\lambda_{2_s4_s}\right) \\[2mm]
\hat\lambda^{\kappa_3}={1\over2}
 \left(\hat\lambda_{1_s2_s}-\hat\lambda_{3_s4_s}\right) \\[2mm]
\hat\lambda^{\kappa_4}={1\over2}
 \left(\hat\lambda_{1_s6_s}-\hat\lambda_{2_s5_s}\right) \\[2mm]
\hat\lambda^{\kappa_5}={1\over2}
 \left(\hat\lambda_{1_s5_s}+\hat\lambda_{2_s6_s}\right) \\[2mm]
\hat\lambda^{\kappa_6}={1\over2}
 \left(\hat\lambda_{3_s6_s}-\hat\lambda_{4_s5_s}\right) \\[2mm]
\hat\lambda^{\kappa_7}={1\over2}
 \left(\hat\lambda_{3_s5_s}+\hat\lambda_{4_s6_s}\right) \\[2mm]
\hat\lambda^{\kappa_8}={1\over2\sqrt{3}}
 \left(\hat\lambda_{1_s2_s}+\hat\lambda_{3_s4_s}
  -2\hat\lambda_{5_s6_s}\right)
\end{array}
\hskip4.3cm
\]
generates space-time $SU(3)$ transformations appearing in four 
dimensions as {\sl color transformations}. 
All remaining independent combinations equal zero.

\subsection*{The General Principle of Equivalence and local geometry}

 In complete analogy with the general relativistic picture of 
gravitational interactions we assume all fundamental interactions to
appear as a consequence of a non trivial parallel transport of space-time 
reference frames. 
 We enforce the General Principle of Equivalence by promoting the 
metric of the rigid space-time to a point dependent metric
$\hat\eta_{AB}+i\hat\varepsilon_{AB}\rightarrow
\hat{g}_{IJ}(\hat{x})+i\hat\omega_{IJ}(\hat{x})$ 
satisfying the maximal compatibility condition
$\hat{\nabla}_K\hat{g}_{IJ}=0$ and $\hat{\nabla}_K\hat{\omega}_{IJ} =0$.
An assignation of  reference frames in every space-time point is again 
a set of maximally independent vector fields $\hat{e}_A^{\ I}$ fulfilling
\[
\hat{e}_A^{\ I}\hat{e}_B^{\ J}\hat{g}_{IJ}=\hat\eta_{AB}
\ \ \ \ \ \ \ \ \ \ \ 
\hat{e}_A^{\ I}\hat{e}_B^{\ J}\hat\omega_{IJ}=\hat\varepsilon_{AB}
\]
Reference frames are so determined up to a point dependent 
$\hat\Lambda(\hat{x})$ transformation
\[
\hat{e}_A^{\ I}\rightarrow \hat\Lambda_A^{\ B}(\hat{x})\hat{e}_B^{\ I}
\]
Comparing physical informations in different space-time points involves 
as usual the connection
\[
\hat\Omega_{K,AB}=
\hat{e}_A^{\ I}\left(\hat\nabla_K\hat{e}_B^{\ J}\right)\hat{g}_{IJ}
\]
belonging this time to the lie algebra $so(1,3)\oplus u(1)\oplus u(1)\oplus 
su(2)\oplus su(3)$ in virtue of the maximal compatibility condition. 
Under a point dependent 
redefinition of reference frames $\hat\Omega_{K,AB}$ transforms 
like an $SO(1,3)\times U(1)\times U(1)\times SU(2)\times SU(3)$
gauge potential
\[
\hat\Omega_{K,AB}\rightarrow 
\hat\Lambda_A^{\ C}\hat\Lambda_B^{\ D}\hat\Omega_{K,CD}+
\hat\Lambda_A^{\ C}\left(\partial_K\hat\Lambda_B^{\ D}\right)
\hat\eta_{CD}  
\]
The components of $\hat\Omega_{K,AB}$  having $K$ as an ordinary space-time
index and $A$,  $B$ both ranging over ordinary or extra space-time have the 
right transformation properties to represent gravitational and 
non-gravitational fundamental gauge fields respectively. 
 To fully appreciate the geometrical meaning of gravitational and
non-gravitational interactions we develop a little further the geometrical
structure.

\vskip0.2cm
 Consider first the case in which a gauge exist such that $\hat\Omega_{K,AB}
\equiv0$. In the absence of forces space-time is flat. It is then possible to 
find global coordinate frames $\hat{x}^A=(x^\alpha,\xi^a)$ such that the metric
structure $\hat{g}_{IJ}+i\hat\omega_{IJ}$ takes everywhere the standard form 
$\hat\eta_{AB}+i\hat\varepsilon_{AB}$.
It is worth stressing that this rigid geometry is very different from a 
fourteen dimensional Minkowskian geometry. We are not allowed to mix real 
and complex directions together, nor complex directions of a different nature. 
 As a flat geometry our generalization is a rather trivial one: the total 
space-time is the direct product of a four dimensional Minkowskian space-time 
times a complex affine-space of real dimension four times a second complex 
affine-space of real dimension six. However --as we show in next sections--
this rigid structure already has very important phenomenological consequences.
It is responsible for the effective four dimensional structure of elementary 
matter fields.
 In physical terms it is better to think of space-time in a different way. 
Picture out the fourteen dimensional flat space-time  as foliated in null four 
dimensional hyper-surfaces of the degenerate skew-symmetric two-form 
$\hat\varepsilon_{AB}$ (space-time looks like the ordinary three dimensional 
Euclidean space foliated in  one dimensional straight field lines of an 
homogeneous magnetic field). On every leave of such a foliation the symmetric 
two-form $\hat\eta_{AB}$ induces an ordinary  Minkowskian structure.
Leaves are trivially embedded in space-time.
Every $1+3$ null hyper-surface is a good candidate for ordinary space-time. 
\begin{figure}[h]
\begin{center}
\epsfysize6cm
\epsffile{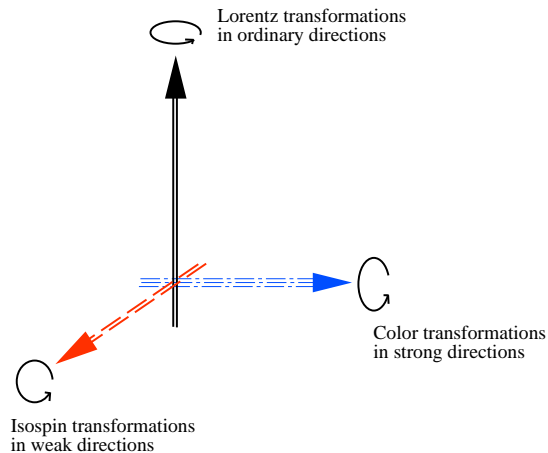}
\end{center}
\caption{A flat space-time may be pictured out as a three dimensional rigid 
space. Axes carry different group of directions. We can only mix directions 
inside axes.}
\end{figure}
\vskip0.2cm
 When fundamental interactions are turned on it is no longer possible to
globally bring the metric structure in the standard form. However, we can 
still picture out the fourteen dimensional space-time as foliated in  
null four dimensional hyper-surfaces of the degenerate two-form 
$\hat\omega_{IJ}(\hat{x})$ (we can maintain the three dimensional analogy 
by replacing the homogeneous magnetic field with an inhomogeneous one).
On every leave of the foliation the symmetric two-form $\hat{g}_{IJ}(\hat{x})$
induces now an $SO(1,3)$ pseudo-Riemannian connection. 
 In the effective four dimensional physics this is identified with
gravitation.
 Every leave is also non trivially embedded in the total space-time.
The twist in space-time of $\hat\omega_{IJ}$ null hyper-surfaces is the 
ingredient making the difference between a product manifold and our
geometrical structure. The effect of such a twist on the effective 
four dimensional physics is that of non-gravitational fundamental
interactions. 
To make the  content of the theory more transparent 
we specialize to coordinates adapted to the foliation. 
Among the various coordinate frames doing the job one choice 
is particularly convenient. As a consequence of the maximal compatibility
condition the two-form $\hat\omega_{IJ}$ is closed  (enforcing
the analogy between $\hat\omega_{IJ}$ and a three dimensional magnetic field). 
Therefore, a classical theorem of Darboux ensures the possibility of   
{\it globally}\footnote{We assume space-time topology to be the one of 
$I\!\!R^{14}$. We also assume the foliation of space-time to be 
regular ignoring problems connected to global integrability. In a general 
topology our consideration are valid in a finite neighborhood of a point.}
finding coordinates in such a way that $\hat\omega_{IJ}\equiv
\hat\varepsilon_{IJ}$. Denoting by $\hat{x}^I=(x^\mu,\xi^i)$ one of such  
Darboux coordinates frames we immediately realize the $x^\mu$ to parameterize  
foliation leaves  and the $\xi^{i_w}$, $\xi^{i_s}$ weak and strong 
directions respectively. Every point $\xi$ in extra space-time 
labels a leave of the foliation. In a Darboux coordinate frame
we  introduce the following parameterization of the Hermitian metric  
\begin{figure}[bt]
\begin{center}
\epsfxsize10cm
\epsffile{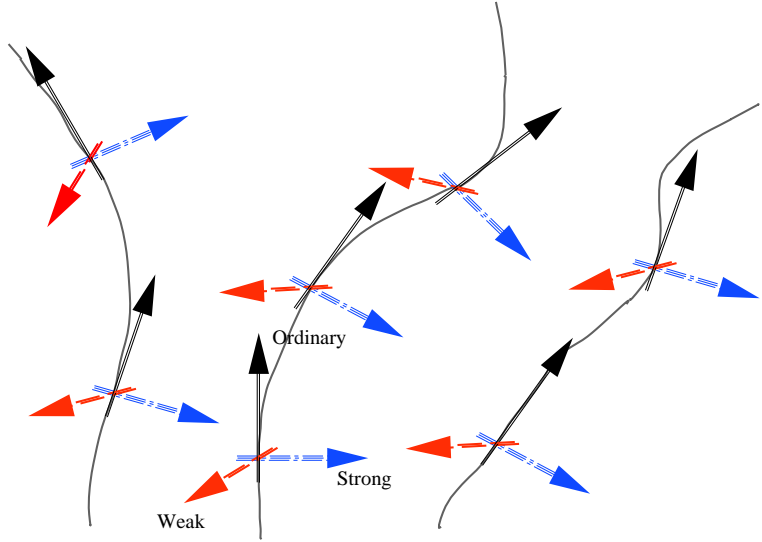}
\end{center}
\caption{A curved space-time is foliated in $1+3$ hyper-surfaces (represented 
here by lines). Ordinary space-time is determined dynamically as one of
these sub-manifolds.}
\end{figure}
\vskip0.2cm
\[
\begin{array}{ccc}
\hat{g}_{IJ}=
\left(
\begin{array}{cc}
 g_{\mu\nu} & g_{\mu\lambda} a_j^\lambda \\[2mm]
 a_i^\kappa g_{\kappa\nu} & g_{ij}+a_i^\kappa a_j^\lambda g_{\kappa\lambda}
\end{array}
\right)
&\ \  \mbox{and} \ \ &
\hat\omega_{IJ}=
\left(
\begin{array}{cc}
 0 & 0 \\[2mm]
 0 & \varepsilon_{ij}
\end{array}
\right)
\end{array}
\]
$g_{\mu\nu}(x,\xi)$ is the pseudo-Riemannian metric induced on the leave 
labeled by $\xi$; $a_i^\mu(x,\xi)$  and $g_{ij}(x,\xi)$ encode  
informations on its extrinsic geometry. 
This parameterization allows to directly relate space-time reference 
frames $\hat{e}_A^{\ I}$ with ordinary $e_\alpha^{\ \mu}$ and extra
$e_a^{\ i}$ reference frames
\[
\hat{e}_A^{\ I}=
\left(
\begin{array}{cc}
 e_\alpha^{\ \mu}& 0 \\[2mm]
-e_a^{\ k}a_k^\mu& e_a^{\ i}
\end{array}
\right)
\hskip6.7cm
\]
Substituting in the general conditions we obtain indeed
\[
\begin{array}{lcl}
e_\alpha^{\ \mu}e_\beta^{\ \nu}g_{\mu\nu}=\eta_{\alpha\beta} & & \\[3mm]

e_a^{\ i}e_b^{\ j}g_{ij}=\delta_{ij} &
\ \ \ \mbox{and} \ \ \               &
e_a^{\ i}e_b^{\ j}\varepsilon_{ij}=\varepsilon_{ab}
\end{array}
\hskip2cm
\]

\vskip0.2cm
The splitting of ordinary and extra space-time coordinates divides  
connection components in six groups: 
$\hat\Omega_{\mu,\alpha\beta}$,
$\hat\Omega_{\mu,\alpha b}$, 
$\hat\Omega_{\mu,ab}$ 
and
$\hat\Omega_{i,\alpha\beta}$, 
$\hat\Omega_{i,\alpha b}$,
$\hat\Omega_{i,ab}$.
The first three describe the properties of the parallel transport of
space-time reference frames in ordinary directions; they completely 
characterize the intrinsic and extrinsic geometry of every leave of the 
foliation as a submanifold of space-time. The last three describe the 
properties of parallel transport of space-time reference frames in 
extra directions; they measure the amount of change in  intrinsic and 
extrinsic geometry when moving from a leave to the next. (In the three
dimensional analogy the former are the equivalent of intrinsic -trivial- 
geometry, curvature and torsion of every field line; the latter are the 
equivalent of vorticity, expansion, shear and similar geometrical properties 
of the foliation of space in terms of field lines.)
If we assume to be constrained to live on a single four dimensional leave 
of the foliation --and in the next sections we show that this is indeed the 
case-- we are only allowed to transport reference frames in ordinary 
directions. Therefore, the geometrical space-time properties we can directly 
perceive in our effective four dimensional physics are the ones described
by $\hat\Omega_{\mu,\alpha\beta}$,
$\hat\Omega_{\mu,\alpha b}$ and $\hat\Omega_{\mu,ab}$. 
\begin{description}
\item{$\hat\Omega_{\mu,\alpha\beta}$} equals the 
{\sl induced connection} on every leave
\[
\hat\Omega_{\mu,\alpha\beta}=\Omega_{\mu,\alpha\beta}
\hskip8.5cm
\]
giving a complete characterization of its intrinsic geometry. The induced 
connection belongs to $so(1,3)$, transforms like a Lorentz gauge potential
and is identified with Einstein's gravitational field. 

\item{$\hat\Omega_{\mu,\alpha b}$} generalizes the notion of curvature of a 
line embedded in the three dimensional Euclidean space. In a higher 
dimensional context it is costume to rewrite it in terms of the 
{\sl extrinsic curvature} (or second fundamental form) 
$\kappa_{\mu\nu a}=e_\mu^{\ \beta}\hat\Omega_{\nu,\beta a}$.
In our geometry $\hat\Omega_{\mu,\alpha b}$ is identically equal to
zero as a consequence of the connection transformation properties
\[
\hat\Omega_{\mu,\alpha b}=e_\alpha^{\ \nu}\kappa_{\mu\nu b}= 0
\hskip7.0cm
\]
Every leave of the foliation is minimally embedded in the total space-time.

\item{$\hat\Omega_{\mu,ab}$} generalizes the notion of torsion of a line 
embedded in the three dimensional Euclidean space. This is perhaps the 
best way of visualizing its geometrical meaning in a higher dimensional 
context. We address these components of the total connection as the 
{\sl extrinsic torsion} of every leave of the foliation. We also introduce 
the special symbol ${\mathcal A}_{\mu,ab}\equiv\hat\Omega_{\mu,ab}$ to denote
it. The extrinsic torsion transforms like a one-form under 
reparameterization   of the leave. It fulfills the conditions  
${\mathcal A}_{\mu,ab}+{\mathcal A}_{\mu,ba}=0$
and                 
${\mathcal A}_{\mu,a}^{\ \ \ c}\varepsilon_{cb}
-{\mathcal A}_{\mu,b}^{\ \ \ c}\varepsilon_{ca}=0$ 
belonging to the Lie algebra $u(1)\oplus u(1)\oplus su(2)\oplus su(3)$.
Under a point dependent redefinition of extra reference frames 
$e_a^{\ i}\rightarrow U_a^{\ b}(x,\xi)e_b^{\ i}$,  
${\mathcal A}_{\mu,ab}$ transforms like a $U(1)\times U(1)\times SU(2)\times 
SU(3)$ gauge connection. 
 In the next section we show that ${\mathcal A}_{\mu,ab}$ effectively couples 
to matter field as a gauge connection. Therefore, we identify the extrinsic
torsion with non-gravitational fundamental interactions. 
 In terms of the parameterization introduced above
\[
{\mathcal A}_{\mu,ab}=
{1\over2}\left(
e_a^{\ i}(\partial_\mu e_b^{\ j})-
(\partial_\mu e_a^{\ i})e_b^{\ j}
\right)g_{ij}
\hskip5.45cm
\]
For later use it is convenient to rearrange the various components
of  ${\mathcal A}_{\mu,ab}$ as follows: the gauge potential associated
to extracharge and hypercharge are chosen as 
${\mathcal A}^x_\mu=
{1\over2w}({\mathcal A}_{\mu,1_w2_w}+{\mathcal A}_{\mu,3_w4_w})+
{1\over2s}({\mathcal A}_{\mu,1_s2_s}+{\mathcal A}_{\mu,3_s4_s}+
           {\mathcal A}_{\mu,5_s6_s})$
and 
${\mathcal A}^y_\mu=
{1\over2w}({\mathcal A}_{\mu,1_w2_w}+{\mathcal A}_{\mu,3_w4_w})-
{1\over2s}({\mathcal A}_{\mu,1_s2_s}+{\mathcal A}_{\mu,3_s4_s}+
           {\mathcal A}_{\mu,5_s6_s})$;
the ones corresponding to isospin 
${\mathcal A}^{\iota_1}_\mu$,
${\mathcal A}^{\iota_2}_\mu$, ${\mathcal A}^{\iota_3}_\mu$ 
and to color
${\mathcal A}^{\kappa_1}_\mu$,${\mathcal A}^{\kappa_2}_\mu$,...,
${\mathcal A}^{\kappa_8}_\mu$ 
are defined according to the notation for infinitesimal generators.
\end{description}
Of the remaining three groups of  connection components, the 
vanishing on symmetry grounds of $\hat\Omega_{i,\alpha b}\equiv0$ guarantees 
the integrability of directions everywhere orthogonal to ordinary space-time
hyper-surfaces. Therefore, we can imagine space-time as further foliated in 
ten dimensional extra hyper-surfaces.  $\hat\Omega_{i,ab}$ is then interpreted
as the connection induced on every such leave and 
$\hat\Omega_{i,\alpha\beta}$ as its extrinsic torsion. \\
 The triviality of the parallel transport of space-time reference frames
is measured, as usual, by the gauge covariant curvature field strength 
\[
\hat{R}_{IJAB}=\partial_I\hat\Omega_{J,AB}-
                \partial_J\hat\Omega_{I,AB}+
                \hat\Omega_{I,A}^{\ \ \ \ \!C}\hat\Omega_{J,CB}-
                \hat\Omega_{I,B}^{\ \ \ \ \!C}\hat\Omega_{J,CA}
\]
The splitting of ordinary and extra space-time coordinates divides 
field strength components in nine groups. On symmetry grounds only 
four of these are non vanishing: $\hat{R}_{\mu\nu\alpha\beta}$,
$\hat{R}_{\mu\nu ab}$ and $\hat{R}_{ij\alpha\beta}$
$\hat{R}_{ijab}$. The ones measuring the triviality of the parallel transport
of reference frames in ordinary directions are 
$\hat{R}_{\mu\nu\alpha\beta}$ and $\hat{R}_{\mu\nu ab}$
\begin{description}
\item{$\hat{R}_{\mu\nu\alpha\beta}$} equals the induced curvature on every 
leave, that is the {\sl gravitational field strength} 
\[
\hat{R}_{\mu\nu\alpha\beta}=R_{\mu\nu\alpha\beta}=
 \partial_\mu\Omega_{\nu,\alpha\beta}
-\partial_\nu\Omega_{\mu,\alpha\beta}+
\Omega_{\mu,\alpha}^{\ \ \ \gamma}\Omega_{\nu,\gamma\beta}-
\Omega_{\nu,\alpha}^{\ \ \ \gamma}\Omega_{\mu,\gamma\beta}
\hskip1.5cm
\]
These equations coincides with the Gauss equations for the embedding of 
ordinary space-time leaves in the total manifold. $R_{\mu\nu\rho\sigma}=
e_\rho^{\ \alpha}e_\sigma^{\ \beta}R_{\mu\nu\alpha\beta}$ is the 
usual Riemann tensor which contraction is used in writing down
Einstein's equations.

\item{$\hat{R}_{\mu\nu ab}$} equals the {\sl non-gravitational field 
strength} associated to the gauge potential ${\cal A}_{\mu,ab}$ 
\[
\hat{R}_{\mu\nu ab}={\cal F}_{\mu\nu ab}=
\partial_\mu{\mathcal A}_{\nu,ab}-
\partial_\nu{\mathcal A}_{\mu,ab}+
{\mathcal A}_{\mu,a}^{\ \ \ c}{\mathcal A}_{\nu,cb}-
{\mathcal A}_{\mu,b}^{\ \ \ c}{\mathcal A}_{\nu,ca}
\hskip1.6cm
\]
Its definition correspond to the Ricci equations for the embedding of 
leaves in the total space-time.
The field strength of hyper, weak and strong forces as well as the one 
associated to the additional extracharge are components of the total 
space-time curvature.
\end{description}
The vanishing on symmetry grounds of $\hat{R}_{\mu\nu\alpha b}\equiv0$ 
is perfectly compatible with the Codazzi-Mainardi equations once we remember 
that the extrinsic curvature is identically equal to zero. 
The remaining two groups of non vanishing components of the total
curvature tensor express analogue properties for the orthogonal foliation: 
$\hat{R}_{ijab}$ equals the curvature field strength induced on every
ten dimensional extra leave while $\hat{R}_{ij\alpha\beta}$ the 
curvature associated to the relative extrinsic torsion.

\subsection*{An extra coupling}

 As a consequence of the maximal compatibility condition the skew-symmetric 
two-form $\hat\omega_{IJ}$ is closed. This induces a twist on the bundle 
of complex fields defined  on the manifold. In more familiar words 
the  principle that all possible coupling have to be included 
in field equations makes $\hat\omega_{IJ}$ to couple to matter  like a
magnetic field (further enforcing the three dimensional analogy). Therefore, 
in all covariant differentiation operators ordinary derivatives $\partial_I$ 
have to be replaced by $\partial_I-il^{-2}\hat\theta_I$; the vector potential 
$\hat\theta_I$ is associated to the two-form $\hat\omega_{IJ}$ in the usual 
way: $\hat\omega_{IJ}=\partial_I\hat\theta_J-\partial_J\hat\theta_I$.
On dimensional grounds we are forced to introduce a {\sl fundamental length} 
$l$ in the theory.
In the absence of other scales we are naturally lead to identify 
$l$ with the Planck length $l\approx 10^{-33}cm$. 
However, all basic feature of elementary matter fields discussed in
this paper only depend on the assumption 
that $l$ is small  ({\it i.e.}~we are considering the theory at energy scales 
much smaller than the one associated to $l$) and not on the particular value 
it takes.
It is worth remarking that $\hat\theta_I$ is not an external 
gauge structure superimposed on the manifold. It is an intrinsic geometrical
object that naturally comes in requiring the structural group of the 
geometry to be --in part-- unitary. \\
 In order to write down field equations for test matter fields 
it is convenient to choose a Darboux coordinate frame $\hat{x}^I
=(x^\mu,\xi^i)$ and partially fix the gauge freedom associated the 
one-form $\hat\theta_I$ by choosing $\hat\theta_I=(0,\theta_i)$. 
Differentiation in ordinary space-time directions is then 
realized by the usual momenta $-i\partial_\mu$. Together with 
the $x^\mu$ these operators close the ordinary space-time Heisenberg 
algebra generating $SO(1,3)$ in the standard way.
Differentiation in weak and strong directions is instead realized by
the operators 
\[
\Pi_i=-il\partial_i-l^{-1}\theta_i \hskip1cm
\]
These ones are no longer a set of mutually commuting operators. They
fulfill commutation relations
\[
\left[\Pi_i,\Pi_j\right]=i\varepsilon_{ij}
\]
where $\varepsilon_{ij}\equiv\hat\varepsilon_{ij}$ is the non-degenerate 
extra part of  $\hat\varepsilon_{IJ}$. The $\Pi_i$ behaves as
five pairs of canonically conjugate operators. For this reason, it is no 
longer convenient to consider the multiplication by $\xi^i$ as conjugate 
to $\Pi_i$. We  rearrange canonical variables by introducing  
the independent set of operators 
\[
\Xi^i=\xi^i+l{\varepsilon}^{ij}\Pi_j \hskip1.45cm
\]
where $\varepsilon_{ik}{\varepsilon}^{jk}=\delta_i^j$.
We can check out immediately that the $\Xi^i$ commute with the $\Pi_i$
and fulfill commutation relations
\[
\left[\Xi^i,\Xi^j\right]=il^2{\varepsilon}^{ji}
\]
The $\Xi^i$ are five more pairs of canonically conjugate operators. 
These complete the representation of Heisenberg algebra in extra directions.
 Infinitesimal generators of the two $U(1)$, of $SU(2)$ and of $SU(3)$ 
--somehow the equivalent of the orbital angular momentum in ordinary 
directions-- are constructed in terms  of $\Pi_i$ as 
\[
L^{ij}={1\over4}(\delta^{ik}{\varepsilon}^{jl}-
                 \delta^{jk}{\varepsilon}^{il})
                 \left\{\Pi_k,\Pi_l\right\}
\]
The {\it extra orbital angular momentum} $L^{ij}$
fulfills the conditions $L_{ij}+L_{ij}=0$
and $L_i^{\ k}\varepsilon_{kj}-L_j^{\ k}\varepsilon_{ki}=0$. Following the 
notation previously introduced we group its components in the generators 
$L^x$, $L^y$ of the two $U(1)$ groups; the generators $L^{\iota_1}$, 
$L^{\iota_2}$, $L^{\iota_3}$ of $SU(2)$; and the generators 
$L^{\kappa_1}$, $L^{\kappa_2}$, ..., $L^{\kappa_8}$ of $SU(3)$.

\newpage
\section{Spinor in a background}

 Unification of fundamental forces in space-time connection 
should be mirrored by unification of elementary matter in a fermionic 
field. The ability of including force and matter in a single 
geometrical frame is a crucial test for the theory. In addition we 
have to justify why our direct perception of the fourteen dimensional 
space-time continuous is reduced to the familiar four real dimensions. 
 To these tasks we consider the free propagation of a test spinor in a 
given fourteen dimensional background. 

\vskip0.2cm
 We start or analysis by drawing an analogy 
with the three dimensional motion of a charged particle in a magnetic field
\cite{Northrop}. 
 The analogy is between the fourteen dimensional space-time and 
the three dimensional space. 
 The background geometry is identified with the Euclidean metric plus 
the magnetic field.
The field is required to have a constant magnitude $1/l$ but is otherwise 
allowed to vary arbitrarily in direction. 
 The test spinor is identified with the charged particle.
 The magnetic field  null direction spans magnetic lines. Therefore, these 
have to be pictured out as ordinary space-time leaves. The two directions 
orthogonal to the magnetic field are instead identified with extra directions. 
 Observe that the analogy is not complete in that the magnetic field is not
required to be parallelly transported by the Euclidean structure and the spin
freedom of the three dimensional particle is disregarded. Nonetheless,
the two systems share the same qualitative dynamical behavior. 
 The three dimensional model is meant as a tool to concretely picture out  
the behavior of the fourteen dimensional spinor in space-time. \\
 Consider first the case of an homogeneous magnetic field (Figure 3a). 
A charged particle moving in such a background performs a rapid rotation of 
radius $l$ in the plane orthogonal to the field while freely propagating along
a straight magnetic line. The stronger the field magnitude the smaller the 
radius of the circle. For very large values of $1/l$ the trajectory of the 
particle is indistinguishable from a straight line. A conservation law protect
the particle from drifting in directions orthogonal to the field so that the 
motion starting in a neighborhood of a line will never drift away. 
The motion of the particle is effectively constrained on a magnetic line. 
\begin{figure}[hbt]
\begin{center}
\epsfxsize15cm
\epsffile{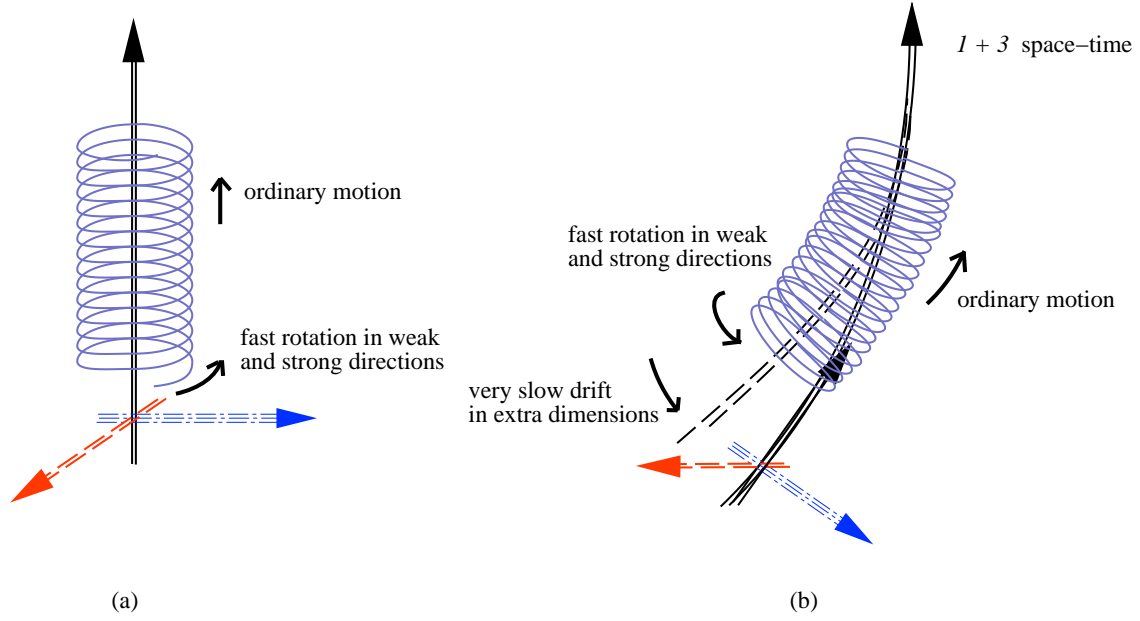}
\end{center}
\caption{The effective motion of a charged particle in a strong magnetic 
background is effectively reduced on a field line. Freedoms divide on
three different energy scales: a very fast rotation, 
the effective motion along the field line and a very slow drift
in normal directions.}
\end{figure}

\noindent
Allow now the field to vary in direction (Figure 3b). 
Magnetic lines are no longer 
straight. However, in the limit of very large values of $1/l$ the motion
of the particle still separate on three different energies scales. 
A consistent fraction of energy (going as $1/l$) is  
stored in a rapid rotation orthogonal to the field direction. 
On a lower energy scale (going as $l^0$) the particle drifts along 
the magnetic line. 
The effective one dimensional motion couples to the magnetic line 
torsion as to a $U(1)$ gauge potential; the effective charge is proportional 
to the angular momentum stored in the fast rotation \cite{Littlejohn}. 
On an even lower energy scale (going as $l^2$) the 
particle drifts in directions normal to the field. In contrast with the 
homogeneous case, the motion starting in a neighborhood of a generic 
magnetic line does not remain there forever. However, if the effective 
Hamiltonian describing the very slow drift in normal directions presents 
an absolute minimum in correspondence of a given magnetic line, whatever 
initial conditions will bring the particle in a neighborhood of that line. 
 The motion of the particle gets effectively constrained on a privileged 
magnetic line.\\
The picture does not change substantially if we switch from classical to 
quantum dynamics \cite{Maraner}. For very large values of $1/l$ 
the wavefunction of the system effectively separates in a part taking 
into account dynamics in normal directions and a part describing the effective 
motion of the particle along the field line. The normal rotational motion 
gets quantized and the particle effectively behaves  as frozen in a
rotational eigenstate --not necessarily the ground state.
Normal rotational eigenstates: are confined in a neighborhood of size 
$l$ of the magnetic line; are labeled by a $U(1)$ quantum number 
corresponding to the angular momentum stored in normal directions; 
are infinitely degenerate as far as 
the very  slow drift in normal directions is negligible.
 In correspondence to every normal eigenstate the particle 
effectively propagates along the field line. The effective quantum motion 
still couples to the field line torsion as to a $U(1)$ gauge field; the
effective charge is quantized and proportional to the  angular 
momentum stored in normal directions.

\vskip0.2cm
 Having this in mind we go back to the original problem. 
By a spinor on a pseudo-Riemannian manifold we mean a field transforming 
according to a spinorial representation of the structural pseudo-orthogonal 
group.
 The natural embedding of the structural group of space-time in $SO(1,13)$ 
allows us a natural definition of space-time spinors.
We introduce Dirac's gamma matrices $\hat\gamma^A$ fulfilling standard
anti-commutation relations
\[
\{\hat\gamma^A,\hat\gamma^B\}=2\hat\eta^{AB} 
\hskip6.5cm
\]
As usual, independent commutators of the gammas generate 
the spin-$1/2$ representation of $SO(1,13)$  
\[
\hat\Sigma^{AB}=-{i\over4}\left[\hat\gamma^A,\hat\gamma^B\right]
\hskip6cm
\]
Among the $\hat\Sigma^{AB}$ we identify ordinary and {\it extra spin angular 
momentum} operators as the linear combinations  generating the subgroup
$SO(1,3)\times U(1)\times U(1)\times SU(2)\times SU(3)$. 
Once more we take advantage of the notation introduced for
infinitesimal generators:  $\hat\Sigma^{\alpha\beta}$ for ordinary
space-time $SO(1,3)$; $\hat\Sigma^x$ and $\hat\Sigma^y$ generating the
two  $U(1)$ subgroups; $\hat\Sigma^{\iota_1}$,  $\hat\Sigma^{\iota_2}$,
$\hat\Sigma^{\iota_3}$ generating $SU(2)$; and 
$\hat\Sigma^{\kappa_1}$, $\hat\Sigma^{\kappa_2}$, ..., $\hat\Sigma^{\kappa_8}$ 
for the generators of $SU(3)$. In fourteen dimensions gamma and sigma matrices
have size $2^7$ by $2^7$ and act on Dirac spinors having $2^7$ components.

\vskip0.2cm
 The free propagation of a Dirac spinor field $\hat\psi(\hat{x})$ in a curved 
background is described by the covariant {\sl matter field equations}
\vskip0.05cm
\[
\left(\hat\gamma^A\hat{e}_A^{\ I}\hat{D}_I-\hat{\sf p}\right) 
 \hat\psi(\hat x)=0
\]
\vskip0.05cm
\noindent
where $\hat{D}_I=-i\partial_I-l^{-2}\hat\theta_I+{1\over2}\hat\Omega_{I,AB}
\hat\Sigma^{AB}$ is the covariant derivative on spinors with extra 
coupling and $\hat{\sf p}$ is a  fourteen dimensional mass term. 
The sign in front of $l^{-2}\hat\theta_I$ is clearly arbitrary. This choice 
corresponds to standard particle physics notations. The sign in front of 
the connection term is fixed by general covariance. Observe  
that since $\hat\Omega_{I,AB}$ belongs to $so(1,3)\oplus u(1)\oplus u(1)
\oplus su(2)\oplus su(3)$ only the linear combinations of $\hat\Sigma^{AB}$
defining ordinary and  extra spin angular momentum appear in the equation. 
This is consistent with the required local covariance.
 As an additional physical proviso we require the integral of 
$\hat\psi^\dagger\hat\psi$ over the whole thirteen dimensional space 
(no time) to be finite. As a consequence not all values of $\hat{\sf p}$ 
yield acceptable solutions. In the following we identify square 
integrable solutions of matter field equations with elementary fermions. \\
Matter field equations are not easy to solve in a generic 
background. On the other hand, we expect on physical grounds that the 
fundamental length $l$ is extremely small  --perhaps of order of the Planck 
scale $l\approx 10^{-33} cm$. This allows us to proceed 
perturbatively in $l$. The qualitative behavior of the system is analogue 
to that of a charged particle in a  strong magnetic background. 
The freedoms of the system divide on three different energy scales 
producing field configurations to squeeze in a neighborhood of a $1+3$ 
hyper-surface. 
To get quantitative we  adopt the following strategy. 
Introduce Darboux coordinates $\hat{x}^I=(x^\mu,\xi^i)$ and partially fix 
extra gauge freedom by choosing $\hat\theta_I=(0,\theta_i)$. 
Recalling that $\xi^i=\Xi^i-l\varepsilon^{ij}\Pi_j$ rewrite 
$\hat\gamma^A\hat{e}_A^{\ I}D_I$ in terms of the canonical operators 
$\Pi_i$, $-i\partial_\mu$, $x^\mu$ and $\Xi^i$ plus ordinary and extra
spin. This produces a natural expansion of field equations in powers of $l$. 
The solution is then constructed by means of ordinary perturbation theory.
 An explicit analysis shows that the leading order term goes as $l^{-1}$
and represents --up to higher order terms-- an harmonic oscillator in the 
canonical variables $\Pi_i$  and extra spin. These are the only 
freedoms active on this energy scale. They are associated to a fast rotation 
of size $l$ in extra directions and carry $U(1)\times U(1)\times SU(2)\times 
SU(3)$ quantum numbers. We will see shortly that these freedom are associated 
to family quantum numbers.
 The next to leading term is of order $l^0$ and  resembles a covariant 
Dirac operator on an effective $1+3$ space-time. The freedoms active on this 
energy scale are the canonical partners $-i\partial_\mu$, $x^\mu$ and 
ordinary spin. These describe effective ordinary motion.
 The remaining freedoms $\Xi^i$ become dynamical only at higher order. 
They are associated to the very slow drift in extra directions.
As long as the effects of such a very slow drift are negligible 
$U(1)\times U(1)\times SU(2)\times SU(3)$ multiplets  are degenerate.
In a while it will become clear that this degeneracy is responsible for 
the phenomenon of families replication. \\
We now focus on the analysis of two simple background configurations 
that allow us to capture the basic feature of the dimensional reduction 
process in a particularly clean way.
The free motion in a flat background --somehow a `special
relativistic' problem-- displays clearly 
the emergence of effective four dimensional chiral fermions, the mechanism 
making them to carry hypercharge, isospin and color quantum numbers 
and the one producing the replication of a basic pattern in different
generations. 
The free motion on an intrinsically flat ordinary 
space-time leave non trivially embedded in space-time allows us to
check how the effective four dimensional motion couples to non-gravitational 
interactions.

\subsection*{Free motion in a flat background} 

In a flat background it is possible and convenient to globally
parallelize reference frames. We take therefore 
$e_\alpha^{\ \mu}\equiv\delta_\alpha^\mu$, $e_a^i\equiv\delta_a^i$
and $a_i^\mu\equiv0$. With this natural choice matter field equations 
take the simple form
\[
\left[{1\over l}\hat\gamma^a\Pi_a
 -i\hat\gamma^\mu\partial_\mu\right]\ \!\hat\psi=\hat{\sf p}\ \!\hat\psi
\]
Ordinary and extra freedoms decouples. The operators $\Xi^i$ are constants of 
motion. The problem is capable of a direct exact solution. 
However, in view of the more complicated case in a curved background we 
proceed by constructing solutions by means of perturbation theory. The 
leading order term  $\hat\gamma^a\Pi_a$ is identified with the solvable 
part of the problem. The remaining terms with the perturbation. 

\vskip0.2cm
\noindent
The {\it leading} order corresponds to the  eigenvalues problem 	 
\[
\hat\gamma^a\Pi_a\ \!\hat\psi_{\underline{\sf p}}=
\ \hat{\sf p}\ \!\hat\psi_{\underline{\sf p}}
\hskip8cm
\]
where the multi-index $\underline{\sf p}$ denotes the set of quantum
numbers necessary to label eigenstates corresponding to the eigenvalue
$\hat{\sf p}$.
In order to construct solutions it is useful to explicitly separate ordinary 
four dimensional and extra ten dimensional spinor indices by  parameterizing 
the $\hat\gamma^A$ in terms of four and ten dimensional gamma matrices. 
Denoting by $\gamma^\mu$, $\mu=0,1,2,3$, ordinary four by four $SO(1,3)$ Dirac 
matrices, by $\gamma^a$, $a=4,5,...,13$, extra thirty-two by 
thirty-two $SO(10)$ gamma matrices and by 
$\gamma_\|=-i\gamma^0\gamma^1\gamma^2\gamma^3$ 
and $\gamma_\bot=i\gamma^{1_w}\gamma^{2_w}...\gamma^{6_s}$ the 
four and ten dimensional chiral matrices, we construct the representation 
\[
\begin{array}{ll}
\hat\gamma^\alpha=\gamma^\alpha\otimes 1\!\!1_{32} &  \alpha=0,1,2,3\\[2mm]
\hat\gamma^a=\gamma_\parallel\otimes\gamma^a & a=4, 5,  ..., 13
\end{array}
\hskip5.5cm
\]
It is readily check that these matrices fulfill the correct anti-commutation 
relations.
In correspondence we rewrite the spinor field as the tensor product of an
ordinary four dimensional spinor times an extra ten dimensional spinor:  
$\hat\psi_{\underline{\sf p}}(\hat{x})=\psi_{\underline{\sf p}}(x)
\otimes\phi_{\underline{\sf p}}(\xi)$. 
Four and ten dimensional components may be chosen as functions of 
ordinary and extra coordinates respectively.
The leading order eigenvalue problem decouples as 
\[
\left\{
\begin{array}{l}
\gamma_\parallel\ \!\psi_{\underline{\sf p}}(x)=
\pm\ \!\psi_{\underline{\sf p}}(x) \\[2mm]
{\cal P}\phi_{\underline{\sf p}}(\xi)=
{\sf p}\ \!\phi_{\underline{\sf p}}(\xi)
\end{array}
\right.
\hskip8cm
\]
where we have introduced the {\sl particle operator} ${\cal P}=\gamma^a\Pi_a$.
For very small values of $l$ extra freedoms are frozen in  ${\cal P}$ 
eigenstates. In the next section we show that particle operator 
eigenstates: are squeezed in a neighborhood of size $l$ of an ordinary $1+3$ 
real hyper-surfaces; are organized in  
$U(1)\times U(1)\times SU(2)\times SU(3)$ multiplets;
every multiplet is repeated in the spectrum
as a consequence of $\Xi^i$ conservation. 
 In correspondence to every particle operator eigenstate the four 
dimensional spinor $\psi_{\underline{\sf p}}(x)$ --which inherit
the multiplet structure and relative degeneracy from the corresponding 
${\cal P}$ eigenstate--
describes the effective four dimensional motion of the system. Leading 
order matter field equations require four dimensional spinors to be 
{\it chiral}. This is a first very important phenomenological prediction of 
the theory. \\
 Recall now that the whole theory is ${\sl C\!\hat{T}\!P}$ invariant. 
Since leading order matter field equations do not depend on bosonic 
ordinary space-time variables, ${\sl C\!\hat{T}\!P}$ decomposes in the 
product of two discrete symmetries which are independently conserved
\[
{\sl C\!\hat{T}\!P}=(\gamma_\|{\sl C\!T\!P}_\|\otimes1\!\!1_{32} ) 
          (\gamma_\| \otimes {\sl C\!P}_\bot)
\hskip5cm
\]
where ${\sl C\!T\!P}_\|$ and ${\sl C\!P}_\bot$ are the four and ten 
dimensional part of the total ${\sl C\!\hat{T}\!P}$ operator
\[
\begin{array}{l}
{\sl C\!T\!P}_\|\psi(x)= - \gamma_\| \ \!\psi^*(-x)  \\[2mm]
{\sl C\!P}_\bot \phi(\xi)= \gamma_\bot \phi^*(-\xi)
\end{array}
\hskip7cm
\]
 The first term in the ${\sl C\!\hat{T}\!P}$ decomposition acts as 
the identity operator on spinor indices. Its role is that of enforcing
${\sl C\!T\!P}_\|$ on four dimensional chiral spinors. 
 The second term relates ordinary and extra spinor indices in a 
non trivial manner. We introduce the special symbol $\hat\Gamma
\equiv\gamma_\| \otimes {\sl C\!P}_\bot$ to denote it. Clearly 
$\hat\Gamma^2=1$. $\hat\Gamma$ is a discrete symmetry dividing
the spectrum in two sectors. Assuming the system in one of these 
sectors the chirality of every four dimensional spinor 
$\psi_{\underline{\sf p}}(x)$ is fixed by the ${\sl C\!P}_\bot$ 
value of the corresponding $\phi_{\underline{\sf p}}(\xi)$. 

\vskip0.2cm
\noindent
 We evaluate {\it next to leading} corrections by means of ordinary 
perturbation theory.
Eigenvalues corrections $\Delta\hat{\sf p}$ and leading order 
eigenfunctions left unspecified by degeneracies are determined by 
diagonalizing the perturbation over degenerate leading order eigenstates.
This amounts to solve the ordinary space-time eigenvalue problem  
$-i\gamma^\mu\partial_\mu \psi_{\underline{\sf p}} 
= \Delta\hat{\sf p}\ \!\psi_{\underline{\sf p}}$.
Eigenvalues corrections $\Delta\hat{\sf p}$ are readily determined by the 
chirality constraint on $\psi_{\underline{\sf p}}(x)$. They all equal zero: 
$\Delta\hat{\sf p}=0$. 
 While degeneracy is not lifted, zero order eigenfunctions are partially 
determined.
In the absence of interactions, four dimensional chiral spinors  
$\psi_{\underline{\sf p}}(x)$ describing the effective motion in ordinary
space-time obey the free massless Dirac 
equation
\[
-i\gamma^\mu\partial_\mu \psi_{\underline{\sf p}} = 0
\hskip8cm
\]
This is consistent with the prescriptions of the Standard Model.

\vskip0.2cm
\noindent
 It is readily checked that $\hat\psi_{\underline{\sf p}}(\hat{x})=
\psi_{\underline{\sf p}}(x)\times\phi_{\underline{\sf p}}(\xi)$ 
are exact solutions of our original problem. No further corrections
have to be evaluated. 

\vskip0.2cm
We have all the necessary ingredients to identify space-time spinors with 
elementary fermions.
 For very small values of $l$ the fourteen dimensional matter field
effectively decouples in the tensor product of 
a {\it four dimensional chiral spinor} $\psi_{\underline{\sf p}}(x)$
times a {\it ten dimensional particle operator eigenspinor}
$\phi_{\underline{\sf p}}(\xi)$. \\
Particle operators eigenspinors are squeezed in a neighborhood of 
an ordinary space-time hyper-surface.  
Particle operators eigenspinors are organized in $U(1)\times U(1)\times
SU(2)\times SU(3)$ multiplets. The whole pattern of multiplets is
repeated many times in the spectrum. Every particle
operator eigenfunction carries a definite value of $C\!P_\bot$.
Given the very large amount of energy necessary to excite transitions
between different particle operator eigenstates the system behaves as
frozen in  extra eigenstates. \\
In correspondence to every particle operator eigenstate the four
dimensional chiral spinor $\psi_{\underline{\sf p}}(x)$ 
propagates freely in ordinary spacetime.
Four dimensional spinors are organized in $U(1)\times U(1)\times
SU(2)\times SU(3)$ multiplets in accordance with the structure of 
their extra partners. The pattern of multiplets is repeated a certain 
number of times.
The chirality of every four dimensional spinor is fixed by the $C\!P_\bot$ 
value of its extra partner. \\
The effective four dimensional {\it qualitative} structure is 
identical to the one of the Standard Model of Elementary Particles. 
 Four dimensional fermions are chiral.
 Multiplet quantum numbers corresponds family quantum numbers. 
 The repetition of a basic pattern of multiplets to the phenomenon of 
families replication. 
 Four dimensional matter field equations in absence of interactions are 
correct. \\
 In order to make the correspondence {\it quantitative} we have
to check two more things.
 First of all, we should investigate how the effective four dimensional 
dynamics couples to a curved background. This allows us to double check
the correct identification of $U(1)\times SU(2)\times SU(3)$ quantum 
numbers with  hypercharge, isospin and color of elementary matter fields 
{\it and} 
the correct identification of extrinsic torsion with hyper, weak and strong 
interactions. 
Second, we have to explicitly evaluate the particle operator spectrum 
checking whether the basic pattern of multiplets corresponds to
the familiar family structure.

\subsection*{Free motion in a minimal curved background}

 In a generic background it is not possible to globally parallelize 
reference frames. Space-time leaves are no longer flat. This yields 
an effective force on particles freely propagating in the 
background. In order to explore how four dimensional 
fermions couple to non-gravitational interactions we assume the existence 
of a privileged {\it intrinsically flat} but {\it extrinsically curved} 
leave to be identified with ordinary space-time. 
 To ensure field configurations to remain in a neighborhood of the
privileged leave we further assume it to correspond to a stationary point 
of reference frames as functions of extra coordinates. 
 Expanding in extra directions we  then obtain the following ansatz 
for reference frames in a neighborhood of the privileged leave: 
$e_\alpha^{\ \mu}=\delta_{\alpha}^{\mu} +{\mathcal O}(l)$,
$e_a^{\ i}=e_a^i(x)+{\mathcal O}(l^2)$ and $a_i^\mu= {\mathcal O}(l)$,
where all the functions are evaluated on the leave. 
This approach has the disadvantage that the extra freedoms 
$\Xi^i$ are assumed from the very beginning to be non-dynamical. 
The method can not be used to evaluate higher order corrections. It is 
nevertheless sufficient for our present task. 
 By explicitly computing connection components we obtain matter field 
equations up to higher order in $l$ as 
\[
\left[{1\over l}\hat\gamma^a e_a^{\ i}(x)\Pi_i
+\hat\gamma^\mu(-i\partial_\mu+{1\over2}{\mathcal A}_{\mu,ab}\hat\Sigma^{ab})
+{\mathcal O}(l)\right] \hat\psi=\hat{\sf p}\ \hat\psi
\]
Ordinary and extra freedoms no longer decouples. 
We can not proceed to a direct solution of the problem. 
On the other hand, we can  still proceed perturbatively.

\vskip0.2cm
\noindent
 The {leading} order term of the operator depends now explicitly 
on ordinary space-time bosonic variables. In order to apply 
our previous analysis a little preparatory work is necessary.
As a first thing we observe that --since extra reference
frames fulfill the identity 
$e_a^{\ i}(x)e_b^{\ j}(x)\varepsilon_{ij}=\varepsilon_{ab}$-- 
the operators 
$e_a^{\ i}(x)\Pi_i$ 
close canonical commutation relations 
$[e_a^{\ i}\Pi_i,e_b^{\ i}\Pi_j]=i\varepsilon_{ab}$.
Therefore, the $x$ dependence of the leading order term is fictitious and 
can be removed by an appropriate redefinition of bosonic canonical variables. 
To this task we look for a purely bosonic unitary transformation ${\sl U}$  
such that 
\[
{\sl U}e_a^{\ i}(x)\Pi_i {\sl U}^{\dagger}=\Pi_a
\hskip6cm
\]
where the $\Pi_a$ are new canonical operators fulfilling $[\Pi_a,\Pi_b]=
i\varepsilon_{ab}$. 
Operating ${\sl U}$ on field equations we bring the leading order term back 
to the simple form $\hat\gamma^a\Pi_a/l$. At the same time we generate a new 
term $-i\hat\gamma^\mu {\sl U}(\partial_\mu {\sl U}^{\dagger})$ 
in the perturbation. 
 In the general case the transformation ${\sl U}$  brings the $\Xi^i$ 
--not dynamical here-- in a new set of canonical variables.
However, we can  insist the new operators to  commute with ordinary 
space-time canonical operators; we require 
$[-i\partial_\mu,\Pi_a]=0$. This condition yields the
equation 
\[
\left[U(\partial_\mu U^{\dagger}),\Pi_a\right]={\cal A}_{\mu,a}^{\ \ \ b}\Pi_b
\hskip7cm
\]
allowing us the determination of the quantity 
${\sl U}(\partial_\mu {\sl U}^{\dagger})$ 
without even knowing the explicit form of ${\sl U}$. 
We redly obtain the solution  
\[
{\sl U}(\partial_\mu {\sl U}^{\dagger})=-{i\over2}{\cal A}_{\mu,ab}L^{ab}
\hskip7.2cm
\]
where $L^{ab}$ are the extra orbital angular momentum operators 
introduced in the previous section. 
This allows us to operate the unitary transformation. 
Denoting by $\hat{J}^{ab}=L^{ab}-\hat\Sigma^{ab}$ 
the {\it extra total angular momentum} operators we obtain the matter
field equations in the form  
\[
\left[{1\over l}\hat\gamma^a \Pi_a
+\hat\gamma^\mu\left(-i\partial_\mu
   -{1\over2}{\mathcal A}_{\mu,ab}(x)\hat{J}^{ab}\right)
+{\mathcal O}(l)\right] \hat\psi=\hat{\sf p}\ \hat\psi
\]
Up to extrinsic torsion and higher order terms these  
resemble matter field equations in a flat background. Observe that 
${\mathcal A}_{\mu,ab}$ couples to ordinary variables like a 
non-gravitational gauge connection. 
Extra indices play the role of gauge indices. 
 In order to bring matter field equations in an even more suggestive form 
we explicitly introduce the two $U(1)$, $SU(2)$ and $SU(3)$ infinitesimal 
generators. Without further delay we address these extra space-time symmetries
as extracharge, hypercharge, isospin and color. 
 Decompose $\hat{J}^{ab}$ in 
$\hat{J}^x=L^x-\hat\Sigma^x$, 
$\hat{J}^y=L^y-\hat\Sigma^y$, ... etc.,  
according to the notation introduced in the previously section. 
Recall the similar decomposition for the extrinsic torsion 
${\mathcal A}_{\mu,ab}$ in ${\mathcal A}^x_\mu$, ${\mathcal A}^y_\mu$, ...
etc..
 Denote by ${\sf X}\equiv\hat{J}^x$ the $U(1)$ {\sl extracharge} generator, 
by ${\sf Y}\equiv\hat{J}^y$ the $U(1)$ {\sl hypercharge} generator, by 
${\sf I}^{\sf i}\equiv\hat{J}^{\iota{\sf i}}$, ${\sf i}=1,2,3$, the $SU(2)$ 
{\sl isospin} generators, and by ${\sf C}^{\sf c}\equiv
\hat{J}^{\kappa{\sf c}}$, 
${\sf c}=1,2, ...,8$, the $SU(3)$ {\sl color} generators. 
Matter field equations take then the form 
\[
\left[{1\over l}\hat\gamma^a\Pi_a+
\hat\gamma^\mu\left(
 -i\partial_\mu 
-{\sf X}{\mathcal A}^x_\mu 
-{\sf Y}{\mathcal A}^y_\mu 
-{\mathcal A}^{\iota{\sf i}}_\mu  {\sf I}^{\sf i}
-{\mathcal A}^{\kappa{\sf c}}_\mu {\sf C}^{\sf c}\right)
+{\mathcal O}(l)
\right]\hat\psi=\hat{\sf p} \ \hat\psi
\]
Ordinary and extra freedom are still coupled. In order to 
effectively decouple them we precede by means of perturbation theory.

\vskip0.2cm
\noindent
The discussion of the {\it leading} order goes exactly as in a flat background.

\vskip0.2cm
\noindent
 The analysis gets a little more subtle when {\it next to leading} corrections 
are evaluated. As before the chirality constraint on  
$\psi_{\underline{\sf p}}(x)$ forces  eigenvalues corrections 
$\Delta\hat{\sf p}$ to vanish identically. 
 On the other hand, the presence of isospin ${\sf I}^{\sf i}$ and color 
${\sf C}^{\sf c}$ operators in the perturbation produces now a 
coupling among extra eigenstates  belonging to a $U(1)\times U(1)\times SU(2)
\times SU(3)$ multiplet.
As a consequence the effective equations of motion for four dimensional 
spinors belonging to a multiplet no longer decouples.
 In order to make the analysis transparent it is useful to explicitly 
introduce indices labeling different states of  $U(1)\times U(1)\times 
SU(2)\times SU(3)$ multiplets. We denote by $\phi_{\underline{\sf p},\rm{a}}
\equiv|\underline{\sf p},{\rm a}\rangle$, ${\rm a}=1,...,n$, particle operator 
eigenstates belonging to a given multiplet of dimension $n$. 
Introduce the matrices 
\[
\begin{array}{ll}
{\sf I}^{\sf i}_{{\sf p},{\rm ab}}=
\langle\underline{\sf p},{\rm a}|
        {\sf I}^{\sf i}
|\underline{\sf p},{\rm b}\rangle & {\sf i}=1,2,3 \\[2mm]
{\sf C}^{\sf c}_{{\sf p},{\rm ab}}=
\langle\underline{\sf p},{\rm a}|
{\sf C}^{\sf c}
|\underline{\sf p},{\rm b}\rangle & {\sf c}=1,2, ..., 8
\end{array}
\hskip7cm
\]
where Roman indices ${\rm a}$ and ${\rm b}$ run over the degeneracy
dimension $n$. It is immediate to check that 
${\sf I}^{\sf i}_{{\sf p},{\rm ab}}$
and 
${\sf C}^{\sf c}_{{\sf p},{\rm ab}}$ 
furnish an $n$ by $n$ representation of $SU(2)\times SU(3)$.
Since ${\sf X}$ and ${\sf Y}$ commute with isospin and color generators, 
the representation carries  definite extracharge and hypercharge values:  
$\langle\underline{\sf p},{\rm a}|{\sf X}|\underline{\sf p},{\rm b}\rangle
={\sf X}_{\sf p}1\!\!1_{\rm ab}$, 
$\langle\underline{\sf p},{\rm a}|{\sf Y}|\underline{\sf p},{\rm b}\rangle 
={\sf Y}_{\sf p}1\!\!1_{\rm ab}$.
 Diagonalizing the perturbation over leading order eigenstates 
amounts to solve the massless Dirac problems for chiral spinors coupled 
to a $U(1)\times U(1)\times SU(2)\times SU(3)$ gauge field
\[
\sum_{\rm b}
\gamma^\mu(-i1\!\!1_{\rm ab}\partial_\mu
           -{\sf X}_{\sf p}1\!\!1_{\rm ab}{\cal A}^x_\mu
           -{\sf Y}_{\sf p}1\!\!1_{\rm ab}{\cal A}^y_\mu
-{\cal A}^{\iota{\sf i}}{\sf I}^{\sf i}_{{\sf p},{\rm ab}}
-{\cal A}^{\kappa{\sf c}}{\sf C}^{\sf c}_{{\sf p},{\rm ab}})
           \psi_{\underline{\sf p},{\rm b}} =0
\]
By recompressing degeneracy indices inside the multi-index 
$\underline{\sf p}$ we bring the equation in the standard form.  
Effective four dimensional matter field equations reads
\[
\gamma^\mu(-i\partial_\mu
           -{\sf X}_{\sf p}{\cal A}^x_\mu
           -{\sf Y}_{\sf p}{\cal A}^y_\mu
-{\cal A}^{\iota{\sf i}}{\sf I}^{\sf i}_{\sf p}
-{\cal A}^{\kappa{\sf c}}{\sf C}^{\sf c}_{\sf p})
           \psi_{\underline{\sf p}} =0
\]
where multiplet and spinor indices are understood. Up to extracharge 
coupling these equations reproduce correctly the prescriptions of 
the Standard Model of Elementary Particle. The identification of 
$U(1)\times SU(2)\times SU(3)$ quantum numbers with hypercharge,
isospin and color as well as the identification of extrinsic torsion
with hyper, weak and strong interactions are therefore consistent.

\section{Fermion quantum numbers}

A catalog of elementary particles is obtained by explicitly analyzing
the spectrum of the particle operator 
\[
{\cal P}=\gamma^a\Pi_a
\]

\subsection*{Symmetries}
The first thing striking us is the simplicity in form of ${\cal P}$. 
The particle operator is obtained as the contraction of 
ten extra bosonic operators $\Pi_a$ fulfilling commutation 
relations $[\Pi_a,\Pi_b]=i\varepsilon_{ab}$
with ten extra fermionic operators $\gamma^a$ fulfilling anti-commutation 
relations $\{\gamma^a,\gamma^b\}=\delta^{ab}$. 
The form is  invariant under the exchange of bosonic and fermionic 
freedoms. Therefore, we expect the presence of a {\it supersymmetry}.
In analogy with ${\cal P}=\gamma^a\delta_a^{\ b}\Pi_b$ we introduce a second 
operator $\bar{\cal P}=\gamma^a\varepsilon_a^{\ b}\Pi_b$ where 
$\varepsilon_a^{\ b}$ 
--somehow the anti-symmetric analogue of $\delta_a^{\ b}$--
has the same form of $\varepsilon_{ab}$ with $w$ and $s$
set equal to one\footnote{It is worth remarking that $\varepsilon_a^{\ b}$
as well as the previously introduced $\varepsilon^{ab}$ are not obtained 
from $\varepsilon_{ab}$ by highering indices with the flat metric.}.
We can readily check that ${\cal P}$ and $\bar{\cal P}$ anticommute while 
their squares both equal the infinitesimal generator of extracharge
transformations ${\sf X}=\delta^{ab}\Pi_a\Pi_b-\varepsilon_{ab}\Sigma^{ab}$.
This is actually the reason for choosing the generators of the two 
$U(1)$ groups as we did. Moreover, the particle operator and its
supersymmetric partner anticommute with the extra chiral gamma $\gamma_\bot$. 
The operators ${\cal P}$, $\bar{\cal P}$,
${\sf X}$ and $\gamma_\bot$ close the $N=2$ superalgebra
\[
\begin{array}{c}
\{{\mathcal P},{\mathcal P}\}=
\{\bar{\mathcal P},\bar{\mathcal P}\}=2{\sf X} \\[3mm]
\{{\mathcal P},\bar{\mathcal P}\}=
\{{\mathcal P},\gamma_\bot\}=
\{\bar{\mathcal P},\gamma_\bot\}=0
\end{array}
\]
As a consequence we can define an operator $S=-i\gamma_\bot\bar{\cal P}$
commuting with ${\cal P}$ that can be used to label particle operator 
eigenstates. The square of $S$ equals the extracharge 
${S}^2={\sf X}$. The relevant information is therefore a 
sign. We define the operator $\varsigma=\mbox{sign}(C\!P_\bot S)$
dividing the spectrum  in the sectors $\varsigma=\pm1$.

\vskip0.2cm
\noindent
A direct computation shows that the particle operator commutes with the 
generators of extracharge, hypercharge, isospin and color transformations.
All together the operators ${\cal P}$, $S$, ${\sf X}$, ${\sf Y}$,
${\sf I}^{\sf i}$,  ${\sf i}=1,2,3$, ${\sf C}^{\sf c}$, ${\sf c}=1,...,8$
close the commutation relations 
\[
\begin{array}{c}
[{\cal P},S]=[{\cal P},{\sf X}]=[{\cal P},{\sf Y}]=
[{\cal P},{\sf I}^{\sf i}]=[{\cal P},{\sf C}^{\sf c}]=0 \\[3mm]
[S,{\sf X}]=[S,{\sf Y}]=[S,{\sf I}^{\sf i}]=[S,{\sf C}^{\sf c}]=0 \\[3mm]
[{\sf X},{\sf Y}]=[{\sf X},{\sf I}^{\sf i}]=[{\sf X},{\sf C}^{\sf c}]=0\\[3mm]
[{\sf Y},{\sf I}^{\sf i}]=[{\sf Y},{\sf C}^{\sf c}]=0 \\[3mm]
[{\sf I}^{\sf i},{\sf C}^{\sf c}]=0
\end{array}
\]
and
\[
[{\sf I}^{\sf i},{\sf I}^{\sf j}]=i\epsilon^{\sf ijk}{\sf I}^{\sf k}
\hskip1.0cm
[{\sf C}^{\sf c},{\sf C}^{\sf d}]=i{\sf f}^{\sf cde}{\sf C}^{\sf e}
\]
where $\epsilon^{\sf ijk}$ and ${\sf f}^{\sf cde}$ are standard 
$su(2)$ and $su(3)$ structure constants. 
Extracharge, hypercharge, isospin and color quantum number can be used to label
particle operator eigenstates. 
Eigenvalues of ${\cal P}$ equal the square root of 
extracharge up to a sign. Choosing to explicitly indicate the value of 
${\sf X}$ among eigenstates labels, only  the sign  
$\pi=\mbox{sign}({\cal P})$ of the particle operator matters.
For every eigenstate we indicate:
the value of $\pi$;
the value of $\varsigma$;
the value of ${\sf X}$; 
the value of ${\sf Y}$; 
the isospin representation ${\sf I}={\mathbf 1},{\mathbf 2},{\mathbf 3},...$
 and the value of its third generator ${\sf I}_3$; 
the color representation ${\sf C}={\mathbf 1},{\mathbf 3},\bar{\mathbf 3},
...$ and the values of its third and eighth generators 
${\sf C}_3$ and ${\sf C}_8$.
Representations of $su(2)$ and $su(3)$ are indicated by their dimension in 
boldface characters as long as such a notation is unambiguous. 
Isospin and color quantum numbers will also be represented collectively 
by the sole representations ${\sf I}$ and ${\sf C}$ when the focus 
is on the multiplet and not on the particular state.

\vskip0.2cm
\noindent
Finally, we should remember that ${\cal P}$, $S$, ${\sf X}$, ${\sf Y}$,
${\sf I}^{\sf i}$,  ${\sf i}=1,2,3$, ${\sf C}^{\sf c}$, ${\sf c}=1,...,8$,
commute with the ten canonical operators controlling the very slow drift in 
extra  directions
\[
[{\cal P},\Xi^a]=
[S,\Xi^a]=
[{\sf X},\Xi^a]=
[{\sf Y},\Xi^a]=
[{\sf I}^{\sf i},\Xi^a]=
[{\sf C}^{\sf c},\Xi^a]=0
\]
where $a=1_w,...,4_w,1_s,...,6_s$.
With the $\Xi^a$ we can construct up to five mutually commuting operators 
labeling different generations of particles. The form of these operators and 
hence of the quantum numbers $\underline{\mathbf G}$ 
labeling 
generations are presumably determined by higher order corrections in 
the perturbative expansion. 

\vskip0.2cm
\noindent
Particle operator eigenstates 
$\phi_{\underline{\sf p}}(\xi)\equiv |\ \!\underline{\sf p}\ \!\rangle$ 
are completely specified by 
\vskip0.2mm
\[
|\ \!\underline{\sf p}\ \!\rangle=
\left|\pm;\pm;{\sf X};{\sf Y};{\sf I}, {\sf I}_3;
{\sf C},{\sf C}_3,{\sf C}_8;\ \underline{\mathbf G}\ \!\right>
\]
\vskip0.2mm
\noindent
where the first sign is refers to $\pi$ and the second to $\varsigma$.
In this paper we do not consider effects of
symmetry breaking. As a consequence we omit generations quantum numbers
among eigenstates labels in the sequel.

\subsection*{Creation and annihilation operators}
 In order to explicitly construct eigenvalues and eigenstates of the 
particle operator it is useful to introduce creation and 
annihilation operators corresponding to bosonic and fermionic extra  
freedoms. In weak directions we define 
\[
\begin{array}{ll}
\left\{ \begin{array}{l}
         a_{1w}         ={1\over\sqrt{2w}}\left(\Pi_{1_w}+i\Pi_{2_w}\right)\cr
        {a_{1w}}^\dagger={1\over\sqrt{2w}}\left(\Pi_{1_w}-i\Pi_{2_w}\right)
        \end{array} \right. 
&
\left\{ \begin{array}{l}
         a_{2w}         ={1\over\sqrt{2w}}\left(\Pi_{3_w}+i\Pi_{4_w}\right)\cr
        {a_{2w}}^\dagger={1\over\sqrt{2w}}\left(\Pi_{3_w}-i\Pi_{4_w}\right)
        \end{array} \right. 
\\[5mm]
\left\{ \begin{array}{l}
         b_{1w}         ={1\over2}\left(\gamma^{1_w}+i\gamma^{2_w}\right)\cr
        {b_{1w}}^\dagger={1\over2}\left(\gamma^{1_w}-i\gamma^{2_w}\right)
        \end{array} \right.
&
\left\{ \begin{array}{l}
         b_{2w}         ={1\over2}\left(\gamma^{3_w}+i\gamma^{4_w}\right)\cr
        {b_{2w}}^\dagger={1\over2}\left(\gamma^{3_w}-i\gamma^{4_w}\right)
        \end{array} \right.
\end{array}
\hskip4.5cm
\]
while in strong directions
\[
\begin{array}{lll}
\left\{ \begin{array}{l}
         a_{1s}         ={1\over\sqrt{2s}}\left(\Pi_{1_s}+i\Pi_{2_s}\right)\cr
        {a_{1s}}^\dagger={1\over\sqrt{2s}}\left(\Pi_{1_s}-i\Pi_{2_s}\right)
        \end{array} \right.
&
\left\{ \begin{array}{l}
         a_{2s}         ={1\over\sqrt{2s}}\left(\Pi_{3_s}+i\Pi_{4_s}\right)\cr
        {a_{2s}}^\dagger={1\over\sqrt{2s}}\left(\Pi_{3_s}-i\Pi_{4_s}\right)
        \end{array} \right. 
&
\left\{ \begin{array}{l}
         a_{3s}         ={1\over\sqrt{2s}}\left(\Pi_{5_s}+i\Pi_{6_s}\right)\cr
        {a_{3s}}^\dagger={1\over\sqrt{2s}}\left(\Pi_{5_s}-i\Pi_{6_s}\right)
        \end{array} \right. 
\\[5mm]
\left\{ \begin{array}{l}
         b_{1s}         ={1\over2}\left(\gamma^{1_s}+i\gamma^{2_s}\right)\cr
        {b_{1s}}^\dagger={1\over2}\left(\gamma^{1_s}-i\gamma^{2_s}\right)
        \end{array} \right.
&
\left\{ \begin{array}{l}
         b_{2s}         ={1\over2}\left(\gamma^{3_s}+i\gamma^{4_s}\right)\cr
        {b_{2s}}^\dagger={1\over2}\left(\gamma^{3_s}-i\gamma^{4_s}\right)
        \end{array} \right.
&
\left\{ \begin{array}{l}
         b_{3s}         ={1\over2}\left(\gamma^{5_s}+i\gamma^{6_s}\right)\cr
        {b_{3s}}^\dagger={1\over2}\left(\gamma^{5_s}-i\gamma^{6_s}\right)
        \end{array} \right.
\end{array}
\]
It is readily checked that $a_{\rm i}$, ${a_{\rm i}}^\dagger$, 
$b_{\rm i}$ and ${b_{\rm i}}^\dagger$, ${\rm i}=1w,2w,1s,2s,3s$ fulfill
standard commutation, anti-commutation relations 
$\left[a_{\rm i},{a_{\rm j}}^\dagger\right]=\delta_{{\rm ij}}$
and 
$\left\{b_{\rm i},{b_{\rm j}}^\dagger\right\}=\delta_{{\rm ij}}$. 
We rewrite the whole algebra of operators in terms of creators and 
annihilators. \\
The particle operator ${\cal P}$ takes the form
\[
\begin{array}{l}
{\mathcal P}=\sqrt{2w}\left( a_{1w}{b_{1w}}^\dagger+{a_{1w}}^\dagger b_{1w}+
              a_{2w}{b_{2w}}^\dagger+{a_{2w}}^\dagger b_{2w}\right)+ 
\\[3mm]
\hskip1cm
+\sqrt{2s}\left( a_{1s}{b_{1s}}^\dagger+{a_{1s}}^\dagger b_{1s}+
              a_{2s}{b_{2s}}^\dagger+{a_{2s}}^\dagger b_{2s}+
              a_{3s}{b_{3s}}^\dagger+{a_{3s}}^\dagger b_{3s}\right)
\end{array}
\]
its supersymmetric partner $\bar{\cal P}$
\[
\begin{array}{l}
\bar{\mathcal P}
=-i\sqrt{2w}\left( a_{1w}{b_{1w}}^\dagger-{a_{1w}}^\dagger b_{1w}+
                  a_{2w}{b_{2w}}^\dagger-{a_{2w}}^\dagger b_{2w}\right)+ 
\\[3mm]
\hskip1cm
-i\sqrt{2s}\left( a_{1s}{b_{1s}}^\dagger-{a_{1s}}^\dagger b_{1s}+
                  a_{2s}{b_{2s}}^\dagger-{a_{2s}}^\dagger b_{2s}+
                  a_{3s}{b_{3s}}^\dagger-{a_{3s}}^\dagger b_{3s}\right)
\end{array}
\]
Even though ${\mathcal P}$ and $\bar{\mathcal P}$ are not diagonal 
is the standard occupation number basis associated to $a_{\rm i}$ and 
$b_{\rm i}$, creators and annihilators remains extremely useful tools in 
constructing the spectrum. Indeed, the extracharge operator ${\sf X}=
{\cal P}^2=\bar{\cal P}^2$ as well as the hypercharge ${\sf Y}$, isospin
${\sf I}_3$ and color ${\sf C}_3$, ${\sf C}_8$ operators labeling 
eigenstates are diagonal in this basis.
For extracharge we obtain 
\[
\begin{array}{l}
{\sf X}=2w\left({a_{1w}}^\dagger a_{1w} +{a_{2w}}^\dagger a_{2w}+
                  {b_{1w}}^\dagger b_{1w}+{b_{2w}}^\dagger b_{2w}\right)
\\[3mm]
\hskip1cm
         +2s\left({a_{1s}}^\dagger a_{1s}+{a_{2s}}^\dagger a_{2s}+
                  {a_{3s}}^\dagger a_{3s}+{b_{1s}}^\dagger b_{1s}+
                  {b_{2s}}^\dagger b_{2s}+{b_{3s}}^\dagger b_{3s}\right)
\end{array}
\hskip1.5cm
\]
while for hypercharge
\[
\begin{array}{l}
{\sf Y}=2w\left({a_{1w}}^\dagger a_{1w} +{a_{2w}}^\dagger a_{2w}+
                  {b_{1w}}^\dagger b_{1w}+{b_{2w}}^\dagger b_{2w}\right)
\\[3mm]
\hskip1cm
         -2s\left({a_{1s}}^\dagger a_{1s}+{a_{2s}}^\dagger a_{2s}+
                  {a_{3s}}^\dagger a_{3s}+{b_{1s}}^\dagger b_{1s}+
                  {b_{2s}}^\dagger b_{2s}+{b_{3s}}^\dagger b_{3s}\right)
\end{array}
\hskip1.5cm
\]
In order to construct eigenstates it is useful to have an explicit expression 
of isospin highering and lowering operators ${\sf I}_\pm={\sf I}_1\pm 
i{\sf I}_2$ other that ${\sf I}_3$. We obtain the isospin algebra in the form
\[
\begin{array}{l}
{\sf I}_+={a_{2w}}^\dagger a_{1w} +{b_{2w}}^\dagger b_{1w}\\[3mm]
{\sf I}_-={a_{1w}}^\dagger a_{2w} +{b_{1w}}^\dagger b_{2w}\\[3mm]
{\sf I}_3={1\over2}\left(
          {a_{2w}}^\dagger a_{2w} -{a_{1w}}^\dagger a_{1w} 
         +{b_{2w}}^\dagger b_{2w} -{b_{1w}}^\dagger b_{1w}\right)
\end{array}
\hskip4.65cm
\]
Highering and lowering operators for the color algebra are also necessary.
We introduce the mnemonic notation 
${\sf C}_{\rightleftharpoons}={\sf C}_1\pm i{\sf C}_2$,
${\sf C}_{\nearrow\!\!\!\swarrow}={\sf C}_4\pm i{\sf C}_5$
and 
${\sf C}_{\nwarrow\!\!\!\searrow}={\sf C}_6\pm i{\sf C}_7$
for the operators moving  states in the $su(3)$ weights plane. 
Arrows indicate 
directions in which states are moved. 
We obtain the color algebra in the form
\[
\begin{array}{l}
{\sf C}_{\rightarrow}={a_{2s}}^\dagger a_{1s} +{b_{2s}}^\dagger b_{1s}\\[3mm]
{\sf C}_{\leftarrow} ={a_{1s}}^\dagger a_{2s} +{b_{1s}}^\dagger b_{2s}\\[3mm]
{\sf C}_{\nwarrow}   ={a_{3s}}^\dagger a_{2s} +{b_{3s}}^\dagger b_{2s}\\[3mm]
{\sf C}_{\searrow}   ={a_{2s}}^\dagger a_{3s} +{b_{2s}}^\dagger b_{3s}\\[3mm]
{\sf C}_{\nearrow}   ={a_{3s}}^\dagger a_{1s} +{b_{3s}}^\dagger b_{1s}\\[3mm]
{\sf C}_{\swarrow}   ={a_{1s}}^\dagger a_{3s} +{b_{1s}}^\dagger b_{3s}\\[3mm]
{\sf C}_3={1\over2}\left(
                      {a_{2s}}^\dagger a_{2s}-{a_{1s}}^\dagger a_{1s}
                     +{b_{2s}}^\dagger b_{2s}-{b_{1s}}^\dagger b_{1s}
                  \right)\\[3mm]
{\sf C}_8={1\over2\sqrt{3}}
                 \left(2{a_{3s}}^\dagger a_{3s}-{a_{2s}}^\dagger a_{2s}
                                               -{a_{1s}}^\dagger a_{1s}
                      +2{b_{3s}}^\dagger b_{3s}-{b_{2s}}^\dagger b_{2s}
                                               -{b_{1s}}^\dagger b_{1s}\right)
\end{array}
\hskip1.25cm
\]
Incidentally we observe that the expression of highering and lowering
operators in terms of creators and annihilators is a particularly happy
one. Indeed, we only have to remember the fundamental $su(2)$ diagram 
$\ \ -_{\!\!\!\!\!\!\!\!\! _1 \ \ _2}\ \ $ to reconstruct ${\sf I}_+$
as the operator annihilating in $1$ and re-creating in $2$ and  ${\sf I}_-$ 
as the one annihilating in $2$ and re-creating in $1$. The same holds for 
$su(3)$. Given our choice of conventions the  realization of the color 
algebra transforms according the $\bar{\mathbf 3}$ representation, so that 
the  fundamental diagram we have to remember is  
$\ \ \bigtriangleup_{\!\!\!\!\!\!\!\!\! _1 \ \ _2}^{^{^{\!\!\!\!\!{3}}}}\ $. 
As an example, it is clear that the operator ${\sf C}_{\nearrow}$ corresponds
to annihilating in $1$ and re-creating in $3$, ${\sf C}_{\searrow}$ to 
annihilating in $3$ and re-creating in $2$, etc.. A different choice of
sign in front of the term $l^{-2}\hat\theta_I$ in matter field equations 
would have produced the representation ${\mathbf 3}$ corresponding to the 
fundamental diagram 
$\ \ \bigtriangledown^{\!\!\!\!\!\!\!\!\! ^2 \ \ ^1}_{_{_{_{_{\!\!\!\!\!{3
}}}}}} \ $.\\
 We have all the necessary ingredients to explicitly construct
particle operator eigenstates 
in terms of occupation number states. As usual, these are obtained  
from the vacuum $|\ \!{\mathbf 0}\ \!\rangle$ by acting with creation
operators
\begin{eqnarray}
|n_{1w},n_{2w},\sigma_{1w},\sigma_{2w};
n_{1s},n_{2s},n_{3s},\sigma_{1s},\sigma_{2s},\sigma_{3s} \ \!\rangle=
\hskip6cm\nonumber\\[3mm]
\hskip1cm
{{{a_{1w}}^\dagger}^{n_{1w}}\over\sqrt{n_{1w}!}}
{{{a_{2w}}^\dagger}^{n_{2w}}\over\sqrt{n_{2w}!}}
{{b_{1w}}^\dagger}^{\sigma_{1w}}
{{b_{2w}}^\dagger}^{\sigma_{2w}}
{{{a_{1s}}^\dagger}^{n_{1s}}\over\sqrt{n_{1s}!}}
{{{a_{2s}}^\dagger}^{n_{2s}}\over\sqrt{n_{2s}!}}
{{{a_{3s}}^\dagger}^{n_{3s}}\over\sqrt{n_{3s}!}}
{{b_{1s}}^\dagger}^{\sigma_{1s}}
{{b_{2s}}^\dagger}^{\sigma_{2s}}
{{b_{3s}}^\dagger}^{\sigma_{3s}}
|\ \!{\mathbf 0}\ \!\rangle
\nonumber
\end{eqnarray}
the integer numbers $n_{\rm i}=0,1,2, ...$ denote 
eigenvalues of bosonic number operators ${a_{\rm i}}^\dagger a_{\rm i}$;
the binary numbers $\sigma_{\rm i}=0,1$ denote eigenvalues of fermionic 
number operators ${b_{\rm i}}^\dagger b_{\rm i}$ 
(clearly ${\rm i}=1w,2w,1s,2s,3s$ and here we are not summing over repeated 
indices). Observe that this is more than a formal expression. Number operators
are explicitly realized as differential operators on square integrable
functions of extra  coordinates and spin. The vacuum 
$|\ \!{\mathbf 0}\ \!\rangle$ as well as all excited states correspond
to real extra wavefunctions having --as harmonic oscillator 
eigenstates-- a typical size $l$. 
This proves the effective confinement of the system in a neighborhood
of size $l$ of an ordinary $1+3$ hyper-surface.
 In order to simplify notation we group weak bosonic occupation
number $n_{w1},n_{w2}$ in $\underline{n_w}$, weak fermionic occupation
numbers $\sigma_{1w},\sigma_{2w}$ in $\underline{\sigma_w}$ ... etc.~.  
We also  write things like 
$|\ \!\underline{2_1},\underline{1_2};
\underline{0_{}},\underline{0_{}}\ \!\rangle$
to denote the state
$|\ \!2,0,0,1;0,0,0,0,0,0\ \!\rangle$. 

\subsection*{Spectrum and particle states}
 The particle operator spectrum is readily determined by recalling that 
${\cal P}^2={\sf X}$. The extracharge ${\sf X}$ consist in the sum of 
two harmonic oscillator corresponding to the frequency $2w$ plus three 
harmonic oscillators corresponding to the frequency $2s$. Therefore, 
the particle operator spectrum is 
\[
{\sf p}=\pm\sqrt{2w{\sf n_w}+2{s\sf n_s}}
\]
with ${\sf n_w},{\sf n_s}=0,1,2...$ . The spectrum is discrete.
In order to explicitly construct eigenstates corresponding to the eigenvalues 
$\pm|{\sf p}|$ we proceed in the following standard manner. 
 Select all occupation number states corresponding to the extracharge 
value ${\sf p}^2$.
 With these states construct all the linear combinations that are 
simultaneously annihilated by ${\sf I}_+$, ${\sf C}_\rightarrow$, 
${\sf C}_\nearrow$ and ${\sf C}_\nwarrow$. 
 Among these restricted number of states select new linear combinations 
having a definite value of ${\cal P}$ and $S$.
 These are the highest weight states of the multiplets corresponding to the 
given eigenvalue.
 Act now on these states with the lowering operators ${\sf I}_-$, 
${\sf C}_\leftarrow$, ${\sf C}_\swarrow$ and ${\sf C}_\searrow$ to 
construct the whole representations. As a check control that the total 
number of ${\cal P}$ eigenstates  coincides with the original number 
of ${\sf X}$ eigenstates. 

\vskip0.2cm
Let us now proceed to the explicit construction  of the first few 
particle operator eigenstates and to the identification of the corresponding 
elementary particles: 
\begin{description}
\item{${\sf p}=0$}\\
To ${\sf X}=0$ corresponds the vacuum state
$|\ \! \underline{0_{}},\underline{0_{}};   
      \underline{0_{}},\underline{0_{}} \ \!\rangle
\equiv|\ \! {\mathbf 0}\ \! \rangle$. 
This is the only particle operator eigenstate with ${\sf p}=0$
\[
|\ \! \ ;\ ;0;0;{\mathbf 1},0;{\mathbf 1},0,0\ \!\rangle=
|\ \! \underline{0_{}},\underline{0_{}};   
      \underline{0_{}},\underline{0_{}} \ \!\rangle
\hskip5.0cm
\]
the signs  $\pi$ and $\varsigma$ are clearly not defined.
The state has hypercharge ${\sf Y}=0$, is an isospin singlet 
${\sf I}={\mathbf 1}$ and a color singlet  ${\sf C}={\mathbf 1}$.
Its $C\!P_\bot$ value is $+1$. The corresponding 
particle is a very non-interactive one. It has the right quantum
numbers to be identified with a right-handed neutrino
\[ 
\psi_{0;0;{\mathbf 1};{\mathbf 1}} \longrightarrow \nu_{eR}
\hskip6.0cm
\]
This identification correspond to assume the system in the 
sector $\hat\Gamma=+1$ fixing the chirality of all particles.

\item{${\sf p}=\pm\sqrt{2w}$}\\
To ${\sf X}=2w$ correspond four occupation number states:
$|\ \!\underline{1_1},\underline{0_{}};
      \underline{0_{}},\underline{0_{}}\ \!\rangle$,
$|\ \!\underline{0_{}},\underline{1_1};
      \underline{0_{}},\underline{0_{}}\ \!\rangle$,
$|\ \!\underline{1_2},\underline{0_{}};
      \underline{0_{}},\underline{0_{}}\ \!\rangle$
and
$|\ \!\underline{0_{}},\underline{1_2};
      \underline{0_{}},\underline{0_{}}\ \!\rangle$.
The last two  are the only ones simultaneously annihilated by 
${\sf I}_+$, ${\sf C}_\rightarrow$, ${\sf C}_\nearrow$ and ${\sf C}_\nwarrow$. 
Explicitly analyzing the action of ${\cal P}$ on them we easily 
construct the particle operator eigenstates
\[
\begin{array}{ll}
\left|\ \!\pm;\pm;2w;2w;{\mathbf 2},\ \ {1\over2};
                      {\mathbf 1},0,0\ \!\right\rangle &=
 {1\over\sqrt{2}}
 \left(
 \left|\ \!\underline{1_2},\underline{0_{}};
      \underline{0_{}},\underline{0_{}}\ \!\right\rangle
  \pm
 \left|\ \!\underline{0_{}},\underline{1_2};
      \underline{0_{}},\underline{0_{}}\ \!\right\rangle
 \right) 
\\[3mm]
\left|\ \!\pm;\pm;2w;2w;{\mathbf 2},-{1\over2};
                      {\mathbf 1},0,0\ \!\right\rangle &=
 {1\over\sqrt{2}}
 \left(
 \left|\ \!\underline{1_1},\underline{0_{}};
      \underline{0_{}},\underline{0_{}}\ \!\right\rangle
  \pm
 \left|\ \!\underline{0_{}},\underline{1_1};
      \underline{0_{}},\underline{0_{}}\ \!\right\rangle
 \right) 
\end{array}
\hskip1.0cm
\]
These  carry hypercharge ${\sf Y}=2w$, are isospin doublets 
${\sf I}={\mathbf 2}$ and color singlets ${\sf C}={\mathbf 1}$.
Their $C\!P_\bot$ value is $-1$.
The corresponding particle is therefore left-handed and has the 
right isospin and color quantum numbers to be identified with a 
leptonic doublet
\[ 
\psi_{2w;2w;{\mathbf 2};{\mathbf 1}} \longrightarrow 
\left(\begin{array}{c}\nu_{eL} \cr e_L\end{array}\right)
\hskip4.7cm
\]
Requiring the doublet hypercharge to be $1/2$ we fix the value 
of the weak parameter $w$ to $1/4$.
This is geometrically very  satisfactory being
the inverse of the number of weak directions.
Choosing  the value of the weak parameter $w$ as $1/4$  fixes the 
weak part of hypercharge of all the remaining particles of the spectrum.

\item{${\sf p}=\pm\sqrt{2s}$}\\
To ${\sf X}=2s$ correspond six occupation number states out of which 
we construct the particle operator eigenstates
\[
\begin{array}{ll}
\left|\ \!\pm;\pm;2s;-2s;{\mathbf 1},0;
      \bar{\mathbf 3},\ \ 0,\ \ {1\over\sqrt{3}}\ \!\right\rangle &=
 {1\over\sqrt{2}}
 \left(
 \left|\ \!\underline{0_{}},\underline{0_{}};
           \underline{1_3 },\underline{0_{}}\ \!\right\rangle
  \pm
 \left|\ \!\underline{0_{}},\underline{0_{}};
           \underline{0_{}},\underline{1_3 }\ \!\right\rangle
 \right) 
\\[3mm]
\left|\ \!\pm;\pm;2s;-2s;{\mathbf 1},0;
      \bar{\mathbf 3},\ \ {1\over2},-{1\over2\sqrt{3}}\ \!\right\rangle &=
 {1\over\sqrt{2}}
 \left(
 \left|\ \!\underline{0_{}},\underline{0_{}};
           \underline{1_2 },\underline{0_{}}\ \!\right\rangle
  \pm
 \left|\ \!\underline{0_{}},\underline{0_{}};
           \underline{0_{}},\underline{1_2 }\ \!\right\rangle
 \right) 
\\[3mm]
\left|\ \!\pm;\pm;2s;-2s;{\mathbf 1},0;
      \bar{\mathbf 3},-{1\over2},-{1\over2\sqrt{3}}\ \!\right\rangle &=
 {1\over\sqrt{2}}
 \left(
 \left|\ \!\underline{0_{}},\underline{0_{}};
           \underline{1_1 },\underline{0_{}}\ \!\right\rangle
  \pm
 \left|\ \!\underline{0_{}},\underline{0_{}};
           \underline{0_{}},\underline{1_1 }\ \!\right\rangle
 \right) 

\end{array}
\hskip1.0cm
\]
These carry hypercharge ${\sf Y}=-2s$, are isospin singlets 
${\sf I}={\mathbf 1}$ and color anti-triplets ${\sf C}=\bar{\mathbf 3}$. 
Their $C\!P_\bot$ value is again $-1$. The corresponding particle  
has therefore the right quantum numbers to be identified with an anti
right-handed quark (which is a left-handed particle). The negative 
value of the hypercharge indicates that the quark is of type 
down\footnote{We should remark that $s$ is positive. A negative value
of $s$ will simply interchange the role of extracharge and hypercharge.} 
\[ 
\psi_{2s;-2s;{\mathbf 1};\bar{\mathbf 3}} \longrightarrow d_R^*
\hskip5.3cm
\]
The identification  requires the particle hypercharge to be $-1/3$ 
fixing the value of the strong parameter $s$ to $1/6$. 
Again, this is  geometrically very satisfactory. As for the weak parameter,
it is the inverse of the number of strong directions. Choosing $s$ to be $1/6$
fixes the strong part of hypercharge of all the remaining particles.
\end{description}

\noindent
The identification of particle operator eigenstates corresponding 
to ${\sf p}=\pm\sqrt{2w}$ and ${\sf p}=\pm\sqrt{2s}$ with a 
neutrino-electron type isospin doublet and with a quark down type color
anti-triplets fix the only two parameters left unspecified by the 
gauge group\footnote{Clearly, the parameters $w$ and $s$ can always be 
rescaled through a redefinition of $l$. What really matters is fixing $w/s$
to 3/2 --the ratio of numbers of strong and weak directions.}. 
Extracharge and hypercharge of all remaining elementary fermions are 
determined by the choice
\[
w={1\over4} \hskip2.5cm s={1\over6}
\]
The correct chirality assignment for the first three states is obtained 
if the fourteen dimensional system is in the sector $\hat\Gamma=+1$. 
The remaining discrete symmetries $\pi$ and $\varsigma$ further divide 
the spectrum in sectors. It is natural to assume 
the system  in a single one of these sectors.
We proceed by analyzing a few more particle states:

\begin{description}
\item{${\sf p}=\pm\sqrt{2w+2s}$}\\
To ${\sf X}=2w+2s$ correspond $24$ occupation number states. Out of these
only 
$|\ \!\underline{1_2},\underline{0_{}};
      \underline{1_3},\underline{0_{}}\ \!\rangle$,
$|\ \!\underline{1_2},\underline{0_{}};
      \underline{0_{}},\underline{1_3}\ \!\rangle$,
$|\ \!\underline{0_{}},\underline{1_2};
      \underline{1_3},\underline{0_{}}\ \!\rangle$
and
$|\ \!\underline{0_{}},\underline{1_2};
      \underline{0_{}},\underline{1_3}\ \!\rangle$
are simultaneously annihilated by the highering operators ${\sf I}_+$,
${\sf C}_\rightarrow$ ${\sf C}_\nearrow$ and ${\sf C}_\nwarrow$. Imposing 
that linear combinations of these are  simultaneous eigenstates of 
${\cal P}$ and $S$ we obtain a linear system of eight equations
for the four coefficients. This is immediately solved yielding the 
highest weight states of four isospin doublets ${\sf I}={\mathbf 2}$, 
color anti-triplets ${\sf C}=\bar{\mathbf 3}$
\[
\begin{array}{l}
\left|\ \!\pm;\pm;2w+2s;2w-2s;{\mathbf 2},{1\over2};
      \bar{\mathbf 3},\ \ 0,\ \ {1\over\sqrt{3}}\ \!\right\rangle=
\\[3mm]
\hskip0.5cm
 {1\over\sqrt{4w+4s}}
\left(\sqrt{2w+2s}
 \left|\ \!\underline{0_{}},\underline{1_2};
           \underline{0_{}},\underline{1_3}\ \!\right\rangle
\pm\sqrt{2w}
 \left|\ \!\underline{1_2},\underline{0_{}};
           \underline{0_{}},\underline{1_3}\ \!\right\rangle
\mp\sqrt{2s}
 \left|\ \!\underline{0_{}},\underline{1_2};
           \underline{1_3},\underline{0_{}}\ \!\right\rangle
 \right) 
\\[3mm]
\hskip0.2cm\mbox{and}
\\[3mm]
\left|\ \!\pm;\mp;2w+2s;2w-2s;{\mathbf 2},{1\over2};
      \bar{\mathbf 3},\ \ 0,\ \ {1\over\sqrt{3}}\ \!\right\rangle=
\\[3mm]
\hskip0.5cm
 {1\over\sqrt{4w+4s}}
\left(
\sqrt{2s}
 \left|\ \!\underline{1_2},\underline{0_{}};
           \underline{0_{}},\underline{1_3}\ \!\right\rangle
+\sqrt{2w}
 \left|\ \!\underline{0_{}},\underline{1_2};
           \underline{1_3},\underline{0_{}}\ \!\right\rangle
\pm\sqrt{2w+2s}
 \left|\ \!\underline{1_2},\underline{0_{}};
           \underline{1_3},\underline{0_{}}\ \!\right\rangle
 \right) 
\end{array}
\]
The remaining multiplets states are easily constructed by means lowering 
operators. 
The multiplets carry hypercharge ${\sf Y}=1/6$. Their
$C\!P_\bot$ value is $+1$, corresponding therefore to right-handed 
particles. These are the correct quantum numbers for anti left-handed 
quark doublets
\[ 
\psi_{2w+2s;2w-2s;{\mathbf 2};\bar{\mathbf 3}} \longrightarrow 
\left(\begin{array}{c}u_L\cr d_L\end{array}\right)^{\!\!*}
\hskip3.9cm
\]
Observe that the four multiplet belongs to different sector
of the theory.

\item{${\sf p}=\pm\sqrt{4w}$}\\
Out of the eight occupation number states corresponding to ${\sf X}=4w$
we construct two ${\cal P}$ eigenstates transforming  
like isospin singlets ${\sf I}={\mathbf 1}$,  color singlets ${\sf C}=
{\mathbf 1}$ and two eigenstates transforming like  isospin triplets
${\sf I}={\mathbf 3}$, color singlet ${\sf C}={\mathbf 1}$. 
Maximal weight states are obtained as 
\[
\begin{array}{l}
\left|\ \!\pm;\pm;4w;4w;{\mathbf 1},0;
                        {\mathbf 1},0,0\ \!\right\rangle = 
\\[3mm]\hskip1.5cm=
 {1\over2}
 \left(
 \left|\ \!\underline{1_1},\underline{1_2};
      \underline{0_{}},\underline{0_{}}\ \!\right\rangle
  -
 \left|\ \!\underline{1_2},\underline{1_1};
      \underline{0_{}},\underline{0_{}}\ \!\right\rangle
  \pm\sqrt{2}
 \left|\ \!\underline{0_{}},\underline{1_11_2};
      \underline{0_{}},\underline{0_{}}\ \!\right\rangle
 \right) 
\\[3mm]
\hskip0.5cm\mbox{and}
\\[3mm]
\left|\ \!\pm;\mp;4w;4w;{\mathbf 3},1;
                        {\mathbf 1},0,0\ \!\right\rangle =
 {1\over\sqrt{2}}
 \left(
 \left|\ \!\underline{2_2},\underline{0_{}};
      \underline{0_{}},\underline{0_{}}\ \!\right\rangle
  \pm
 \left|\ \!\underline{1_2},\underline{1_2};
      \underline{0_{}},\underline{0_{}}\ \!\right\rangle
 \right) 
\end{array}
\hskip1cm
\]
These carry hypercharge ${\sf Y}=1$ and correspond to 
$C\!P_\bot=+1$. The singlet states have the correct quantum 
numbers to be identified with right-handed leptons of electronic type
\[ 
\psi_{4w;4w;{\mathbf 1}; {\mathbf 1}} \longrightarrow e_R
\hskip5.5cm
\]
The identification of isospin triplets is more puzzling. However, we
should remark that the four multiplets belongs to different sectors 
of the theory.

\item{${\sf p}=\pm\sqrt{4s}$}\\
To ${\sf X}=4s$ correspond eighteen occupation number states out of which 
we construct two ${\cal P}$ eigenstates transforming like isospin singlets
${\sf I}={\mathbf 1}$, color triplet ${\sf C}={\mathbf 3}$
and two eigenstates transforming like isospin singlets ${\sf I}={\mathbf 1}$
and according the ${\sf C}=\bar{\mathbf 6}$ color representation. 
Once more we display the explicit form of highest weight eigenstates
\[
\begin{array}{l}
\left|\ \!\pm;\pm;4s;-4s;{\mathbf 1},0;
                   {\mathbf 3},{1\over2},{1\over2\sqrt{3}}\ \!\right\rangle = 
\\[3mm]\hskip1.5cm=
 {1\over2}
 \left(
 \left|\ \!\underline{0_{}},\underline{0_{}};
           \underline{1_2},\underline{1_3}\ \!\right\rangle
  -
 \left|\ \!\underline{0_{}},\underline{0_{}};
           \underline{1_3},\underline{1_2}\ \!\right\rangle
  \pm\sqrt{2}
 \left|\ \!\underline{0_{}},\underline{0_{}};
      \underline{0_{}},\underline{1_21_3}\ \!\right\rangle
 \right) 
\\[3mm]
\hskip0.5cm\mbox{and}
\\[3mm]
\left|\ \!\pm;\mp;4s;-4s;{\mathbf 1},0;
                         \bar{\mathbf 6},0,{2\over\sqrt{3}}\ \!\right\rangle =
 {1\over\sqrt{2}}
 \left(
 \left|\ \!\underline{0_{}},\underline{0_{}};
           \underline{2_3},\underline{0_{}}\ \!\right\rangle
  \pm
 \left|\ \!\underline{0_{}},\underline{0_{}};
           \underline{1_3},\underline{1_3}\ \!\right\rangle
 \right) 
\end{array}
\hskip1cm
\]
All these multiplets carry hypercharge ${\sf Y}=-2/3$ and correspond to 
the $C\!P_\bot$ value $+1$. The color triplets have therefore the correct 
quantum numbers to be identified with  right-handed quarks of type
up 
\[ 
\psi_{4s;-4s; {\mathbf 1}; {\mathbf 3}} \longrightarrow u_R
\hskip5.3cm
\]
Puzzling is the identification of the multiplets transforming 
according the $\bar{\mathbf 6}$ color representation. Again, we 
remark that the four multiplets belong to different sectors of the 
theory.
\end{description}
As a consequence of charge conjugation symmetry the sectors $\pi=+$ and 
$\pi=-$ have exactly the same structure. 
Therefore, we can restrict attention to one of the two. 
In Table 1 we summarize the content in multiplet of the sector 
$\pi=+$ of the particle states investigated so far. 
The sector $\varsigma=+$ reproduces correctly the structure of a family.
\begin{table}[t]
\[
\begin{array}{||c|ccc|cccccc|cl|c||}
\vdots&&&&&\vdots&&&&&\ && \\[3mm] 
&&&&&&&&&&&&\\
&&-&&&1&1&{\mathbf3}&{\mathbf 1}& &&\ \ \ \ {\sl no}&   \\[3mm]
5^{th}&&+&&&1&1&{\mathbf1}&{\mathbf 1}&+&& 
 \ \  ({\mathbf 1}, {\mathbf 1})_{+1}^+ &  e_R 
\\[5mm]\cline{2-10}
&&&&&&&&&&&&\\
&&-&&&{5\over6}&{1\over6}&{\mathbf2}&\bar{\mathbf 3}& &&\ \ \ \ {\sl no}&
\\[3mm]
4^{th}&&+&&&{5\over6}&{1\over6}&{\mathbf2}&\bar{\mathbf3}&+&&
\ \ ({\mathbf 2}, \bar{\mathbf 3})_{+1/6}^+ 
&\left(\begin{array}{r} u_L \cr d_L \end{array}\right)^{\!\!*} 
\\[5mm]\cline{2-10}
&&&&&&&&&&&&\\
&&-&&&{2\over3}&-{2\over3}&{\mathbf1}&\bar{\mathbf6}& &&\ \ \ \ {\sl no}&
\\[3mm]    
3^{rd}&&+&&&{2\over3}&-{2\over3}&{\mathbf1}&{\mathbf3}&+&&
\ \ ({\mathbf 1}, {\mathbf 3})_{-2/3}^+ 
& u_R
\\[5mm]\cline{2-10}
&&&&&&&&&&&&\\
2^{nd}&&+&&&{1\over2}&{1\over2}&{\mathbf2}&{\mathbf1}&-&&
\ \ ({\mathbf 2}, {\mathbf 1})_{+1/2}^-& 
\left(\begin{array}{r}\nu_{eL}\cr e_L\end{array}\right)
\\[5mm] \cline{2-10}
&&&&&&&&&&&&\\
1^{st}&&+&&&{1\over3}&-{1\over3}&{\mathbf1}&\bar{\mathbf3}&-&&
\ \ ({\mathbf 1}, \bar{\mathbf 3})_{-1/3}^- \ \ \ 
& d_R^*
\\[5mm]\cline{2-10}
&&&&&&&&&&&&\\
ground&&+&&&0 & 0 &{\mathbf1}&{\mathbf1}&+&&
 \ \ ({\mathbf 1}, {\mathbf 1})_{\ 0}^+   & \nu_{eR}
\\[5mm]\cline{1-10}\cline{13-13}
&&&&&&&&&&&&\\
{\sl state}&&\varsigma&&&{\sf X}&{\sf Y}&{\sf I}&{\sf C}&{\sl C\!P}_\bot &&
    & {\sl particle}                    
\\[5mm]
\hline\hline
\end{array}
\]
\caption{The bottom of the particle operator spectrum (sector $\pi=+$). 
The content in multiplets of the sub-sector $\varsigma=+$ reproduces 
correctly the family structure.} 
\end{table}
Beyond the family structure, the theory predicts the existence of infinite 
many other elementary fermions corresponding to higher excited states.
Moreover, even if we labeled ${\cal P}$ eigenstates with the names of first 
generation particles $e$, $\nu_e$, $d$, $u$, we should keep in mind that 
because of $\Xi^a$ degeneracy this pattern is repeated in the second generation
$\mu$, $\nu_\mu$, $s$, $c$, in the third generation $\tau$, $\nu_\tau$,
$b$, $t$, and possibly in many others\footnote{The number of generations  is
proportional to the extra-volume and is not necessarily infinite.}. 
There is no way of distinguishing
an electron from a muon or a tau before electro-weak symmetry breaking.
 It is not unreasonable to expect that symmetry breaking is 
produced by the coupling with a fourteen dimensional scalar field or 
simply by higher order terms in the expansion of matter field equations
in a curved background. 
In any case --whatever produces the symmetry breaking-- the comprehension 
of the correct mechanism of mass generation will make the prediction of
new particles and generations in a severe test for the theory. \\
 Let us now focus on Table 1. Is it really natural to 
identify its content with a family of elementary fermions? 
As a matter of fact the assignment of hypercharge, isospin, color and 
chirality of every particle is correct.
A definitive answer to the question can only be given through the exhibition 
of a natural  symmetry breaking mechanism. However, it is not unreasonable 
to assume that a particle could possibly acquire a mass in reason of the 
average value of the extracharge --and perhaps hypercharge and complexity of 
isospin and color representations-- of its right and left components. 
If this were the case, the quark down would be lighter than the quark up and 
--even worst-- than the 
electron. Moreover, the fact that the right-handed components of the 
quark down and the left-handed components of the quark doublet appear
in the spectrum through their anti-particles is not particularly
appealing. To have a deeper insight into the problem we analyze
the content of higher particle operator eigenstates. This can be done
in very general terms. In the discussion of generic particle operator 
eigenstates it is better to indicate isospin and color representations by the 
total $su(2)$ isospin and by the two non negative integer labeling $su(3)$
representations. We first consider states in which only
weak modes are excited. These correspond to eigenvalues {${\sf p}=
\pm\sqrt{2w{\sf n_w}}$ where ${\sf n_w}$ is any non negative integer
\begin{description}
\item{${\sf p}=\pm\sqrt{2w{\sf n_w}}$}\\
By counting the possible way of distributing ${\sf n_w}$ quanta 
in two bosonic plus two fermionic harmonic oscillators,
we realize that to the extracharge value ${\sf X}=2w{\sf n_w}$ correspond 
$4{\sf n_w}$ occupation number states. Out of these we easily select the 
ones simultaneously annihilated by ${\sf I}_+$, ${\sf C}_\rightarrow$, 
${\sf C}_\nearrow$, ${\sf C}_\nwarrow$ and rearrange them in ${\cal P}$,
$S$ eigenstates. We obtain the maximal weight states
\[
\!\!\!
\begin{array}{l}
\left|\ \!\pm;\mp(-)^{\sf n_w};2w{\sf n_w};2w{\sf n_w};
    {{\sf n_w}\over2},{{\sf n_w}\over2};
                   (0,0),0,0\ \!\right\rangle= \\[3mm]\hskip7cm
 {1\over\sqrt{2}}
 \left(
 \left|\ \!\underline{{\sf n_w}_2},\underline{0_{}};
      \underline{0_{}},\underline{0_{}}\ \!\right\rangle
  \pm
 \left|\ \!\underline{({\sf n_w}\!-\!1)_2},\underline{1_2};
      \underline{0_{}},\underline{0_{}}\ \!\right\rangle
 \right) 
\\[3mm]
\hskip0.4cm\mbox{and}
\\[3mm]
\left|\ \!\pm;\pm(-)^{\sf n_w};2w{\sf n_w};2w{\sf n_w};
    {{\sf n_w}\!-\!2\over2},{{\sf n_w}\!-\!2\over2};
                   (0,0),0,0\ \!\right\rangle = 
 {1\over\sqrt{2{\sf n_w}}}
 \left(
 \left|\ \!\underline{1_1({\sf n_w}\!-\!2)_2},\underline{1_2};
      \underline{0_{}},\underline{0_{}}\ \!\right\rangle\right.+
\\[3mm]\hskip4.2cm
\left.
-\sqrt{{\sf n_w}\!-\!1}
 \left|\ \!\underline{({\sf n_w}\!-\!1)_2},\underline{1_1};
      \underline{0_{}},\underline{0_{}}\ \!\right\rangle
 \sqrt{\sf n_w}
 \left|\ \!\underline{({\sf n_w}\!-\!2)_2},\underline{1_11_2};
      \underline{0_{}},\underline{0_{}}\ \!\right\rangle
 \right) 
\end{array}
\]
corresponding to two representation of isospin ${\sf I}={{\sf n_w}\over2}$ 
and two representation of isospin ${\sf I}={{\sf n_w}\!-\!2\over2}$; color
quantum numbers are singlets ${\sf C}=(0,0)$. The total number of 
states contained in the four representations is correctly of $4{\sf n_w}$. 
The particle operator level ${\sf p}=\sqrt{2w{\sf n_w}}$ contains therefore 
the multiplets 
\[ 
\left({{\sf n_w}\over2},(0,0)\right)  \oplus 
\left({{\sf n_w}\!-\!2\over2},(0,0)\right) 
\]
where the second irreducible representation appears only for ${\sf n_w}\geq2$.
All states carry hypercharge ${\sf Y}=2w{\sf n_w}$ and correspond to the
$C\!P_\bot$ value $(-1)^{\sf n_w}$.
\end{description}
Next, we consider particle operator eigenstates in which only strong 
modes are excited. These correspond to eigenvalues 
${\sf p}=\pm\sqrt{2s{\sf n_s}}$ with ${\sf n_s}$ any non negative integer
\begin{description}
\item{${\sf p}=\pm\sqrt{2s{\sf n_s}}$}\\
We count again all possible ways of distributing ${\sf n_s}$ quanta in 
three bosonic plus three fermionic harmonic oscillators. 
In this way we find $4{\sf n_s}^2+2$ occupation number 
states corresponding to the extracharge value ${\sf X}=2s{\sf n_s}$.
Maximal weight states are constructed in analogy with the with the 
previous case. We obtain
\[
\begin{array}{l}
\left|\ \!\pm;\mp(-)^{\sf n_s};2s{\sf n_s};-2s{\sf n_s};0,0;
                   (0,{\sf n_s}),0,{{\sf n_s}\over\sqrt{3}}\ \!\right\rangle=
 \\[3mm]\hskip7cm 
 {1\over\sqrt{2}}
 \left(
 \left|\ \!\underline{0_{}},\underline{0_{}};
       \underline{{\sf n_s}_3},\underline{0_{}}\ \!\right\rangle
  \pm
 \left|\ \!\underline{0_{}},\underline{0_{}};
       \underline{({\sf n_s}\!-\!1)_3},\underline{1_3}\ \!\right\rangle
 \right) 
\\[5mm]
\left|\ \!\pm;\pm(-)^{\sf n_s};2s{\sf n_s};-2s{\sf n_s};0,0;
 (1,{\sf n_s}\!-\!2),
                {1\over2},{2{\sf n_s}\!-\!3\over2\sqrt{3}}\ \!\right\rangle= 
 {1\over\sqrt{2{\sf n_s}}}
 \left(
 \left|\ \!\underline{0_{}},\underline{0_{}};
      \underline{1_2({\sf n_s}\!-\!2)_3},\underline{1_3}\ \!\right\rangle
 \right. +
\\[3mm]
\hskip3.8cm\left.
-\sqrt{{\sf n_s}\!-\!1}
 \left|\ \!\underline{0_{}},\underline{0_{}};
      \underline{({\sf n_s}\!-\!1)_3},\underline{1_2}\ \!\right\rangle
\pm\sqrt{\sf n_s}
 \left|\ \!\underline{0_{}},\underline{0_{}};
      \underline{({\sf n_w}\!-\!2)_3},\underline{1_21_3}\ \!\right\rangle
 \right) 
\\[3mm]
\hskip0.4cm\mbox{and}
\\[3mm]
\left|\ \!\pm;\mp(-)^{\sf n_s};2s{\sf n_s};-2s{\sf n_s};0,0;
(0,{\sf n_s}\!-\!3),0,{{\sf n_s}\!-\!3\over\sqrt{3}}\ \!\right\rangle=
\\[3mm]
\hskip3.0cm
{1\over\sqrt{2{\sf n_s}}}
 \left(
 \left|\ \!\underline{0_{}},\underline{0_{}};
      \underline{1_1({\sf n_s}\!-\!3)_3},\underline{1_21_3}\ \!\right\rangle+
-\left|\ \!\underline{0_{}},\underline{0_{}};
      \underline{1_2({\sf n_s}\!-\!3)_3},\underline{1_11_3}\ \!\right\rangle+
 \right.
\\[3mm]
\hskip3.0cm
 \left.
+ \sqrt{{\sf n_s}\!-\!2}
\left|\ \!\underline{0_{}},\underline{0_{}};
      \underline{({\sf n_s}\!-\!2)_3},\underline{1_11_2}\ \!\right\rangle
\pm\sqrt{\sf n_s}
 \left|\ \!\underline{0_{}},\underline{0_{}};
      \underline{({\sf n_s}\!-\!3)_3},\underline{1_11_21_3}\ \!\right\rangle
 \right) 
\end{array}
\]
corresponding to isospin singlets ${\sf I}=0$ and to the color representations 
${\sf C}=(0,{\sf n_s})$, ${\sf C}=(1,{\sf n_s}\!-\!2)$ and 
${\sf C}=(0,{\sf n_s}\!-\!3)$ respectively. The total number of states 
contained in the six irreducible representations is correctly of 
$4{\sf n_s}^2+2$.
The content in multiplets of the level ${\sf p}=\sqrt{2s{\sf n_s}}$ 
is therefore
\[ 
\left(0,(0,{\sf n_s})\right)\oplus 
\left(0,(1,{\sf n_s}\!-\!2)\right)\oplus
\left(0,(0,{\sf n_s}\!-\!3)\right)
\]
where the second and third irreducible representations appear only 
for ${\sf n_s}\geq2$ and ${\sf n_s}\geq3$ respectively. All states 
carry hypercharge ${\sf Y}=-2s{\sf n_s}$ and correspond to the 
$C\!P_\bot$ value $(-1)^{\sf n_s}$.
\end{description}
Out of these two particular cases, we can reconstruct the contents
in multiplet of a generic particle operator eigenstate 
\begin{description}
\item{${\sf p}=\pm\sqrt{2w{\sf n_w}+2s{\sf n_s}}$}\\
The representations content of the particle operator state
containing ${\sf n_w}$ weak oscillation quanta and ${\sf n_s}$ strong 
oscillation quanta coincides with the direct product of the representations
contents of the particle operator states corresponding to 
${\sf p}=\pm\sqrt{2w{\sf n_w}}$ and 
${\sf p}=\pm\sqrt{2s{\sf n_s}}$. 
In the level ${\sf p}=\sqrt{2w{\sf n_w}+2s{\sf n_s}}$ we find 
twice the representation
\begin{eqnarray}{}
\left({{\sf n_w}\over2},(0,{\sf n_s})\right)\oplus 
\left({{\sf n_w}\over2},(1,{\sf n_s}\!-\!2)\right)\oplus
\left({{\sf n_w}\over2},(0,{\sf n_s}\!-\!3)\right)\oplus
\hskip3.0cm
\nonumber
\\[3mm]
\oplus
\left({{\sf n_w}\!-\!2\over2},(0,{\sf n_s})\right)\oplus 
\left({{\sf n_w}\!-\!2\over2},(1,{\sf n_s}\!-\!2)\right)\oplus
\left({{\sf n_w}\!-\!2\over2},(0,{\sf n_s}\!-\!3)\right)
\nonumber
\end{eqnarray}
where the second irreducible representation in the direct sum appears only  
for ${\sf n_s}\geq2$, the third one for ${\sf n_s}\geq3$, the fourth one for 
${\sf n_w}\geq2$, the fifth one for ${\sf n_w}\geq2$ and ${\sf n_s}\geq2$ and 
the sixth one only for ${\sf n_w}\geq2$ and ${\sf n_s}\geq3$. 
One copy of the total representation correspond to the sector $\varsigma=+$
the other one to the sector $\varsigma=-$.
Maximal weight states corresponding to all irreducible representations 
are easily constructed in terms of the states given above.
All eigenstates carry hypercharge ${\sf Y}=2w{\sf n_w}-2s{\sf n_s}$
and correspond to the $C\!P_\bot$ value $(-1)^{\sf n_w+n_s}$.
\end{description}
\begin{table}[t]
\[
\begin{array}{||c|ccccccccccccccccccccccccccc}
\vdots  &&&&&&&&&&&&&& &&&&&&&&&&&&&\\[0.3mm]
        &&&&&&&?&&&&&&& &&&&&?&&&&&&&&\\[0.3mm]
2        &&
         ?
         &&&&&&&&&&&&\!\!\!\!
         \nu_{eR}^*
         \!\!\!\!&&&&&&&&&&&&
         ?&\\[0.3mm]
11\over6 &&&&&&&&&
         ?
         &&&&&&&&&&&&
         ?
         &&&&&&\\[0.3mm]
5\over3  &&&&
         ?
         &&&&&&&&&&&&\!\!\!\!
         {\mathbf d_R}
         \!\!\!\!&&&&&&&&&&&\\[-2.5mm]
3\over2  &&&&&&&&&&&\!\!\!\!\!\!\!\!\!\!
         \left(\begin{array}{r} e_L \cr \!\nu_{eL}\!\end{array}\right)^{\!\!*}
         \!\!\!\!\!\!\!\!\!\!&&&&&&&&&&&&
         ?
         &&&&\\[-2.5mm]
4\over3  &&&&&&
         ?
         &&&&&&&&&&&&\!\!\!\!
         u_R^*
         \!\!\!\!&&&&&&&&&\\[-2.5mm]
7\over6  &&&&&&&&&&&&&\!\!\!\!\!\!\!\!\!\!
         \left(\begin{array}{r}{\mathbf u_L}\cr {\mathbf d_L}\end{array}\right)
         \!\!\!\!\!\!\!\!\!\!&&&&&&&&&&&&&&\\[-2.5mm]
1        &&&&&&&&\!\!\!\!
         e_R^*
         \!\!\!\!&&&&&&&&&&&&\!\!\!\!
         {\mathbf e_R}
         \!\!\!\!&&&&&&&\\[-2.5mm]
5\over6  &&&&&&&&&&&&&&&\!\!\!\!\!\!\!\!\!\!
         \left(\begin{array}{r}d_L\cr u_L\end{array}\right)^{\!\!*}
         \!\!\!\!\!\!\!\!\!\!&&&&&&&&&&&&\\[-2.5mm]
2\over3  &&&&&&&&&&\!\!\!\!
         {\mathbf u_R}
         \!\!\!\!&&&&&&&&&&&&&&&&&\\[-2.5mm]
1\over2  &&&&&&&&&&&&&&&&&\!\!\!\!\!\!\!\!\!\!
         \left(\begin{array}{r}\!{\mathbf \nu\!\!\!\nu_{eL}}\!\cr 
         {\mathbf e_L}\end{array}\right)
         \!\!\!\!\!\!\!\!\!\!&&&&&&&&&&\\[-2.5mm]
1\over3  &&&&&&&&&&&&\!\!\!\!
         d_R^*
         \!\!\!\!&&&&&&&&&&&&&&&\\[0.3mm]
         &&&&&&&&&&&&&& &&&&&&&&&&&&&\\
0        &&&&&&&&&&&&&&\!\!\!\!
         {\mathbf \nu\!\!\!\nu_{eR}}
         \!\!\!\!&&&&&&&&&&&&&\\
         &&&&&&&&&&&&&& &&&&&&&&&&&&&\\
         &&&&&&&&&&&&&& &&&&&&&&&&&&&\\
\hline
         &&&&&&&&&&&&&& &&&&&&&&&&&&&\\
{\sf X}/{\sf Y}&&&&&&\!\!\!...\!\!\!&&\!\!-1\!\!&\!\!-{5\over6}\!\!&
\!\!-{2\over3}\!\!&\!\!-{1\over2}\!\!&\!\!-{1\over3}\!\!&\!\!-{1\over6}\!\!&
0&\!\!+{1\over6}\!\!&\!\!+{1\over3}\!\!&\!\!+{1\over2}\!\!&\!\!+{2\over3}\!\!&
\!\!+{5\over6}\!\!&\!\!+1\!\!&&\!\!\!...\!\!\!&&&&&\\
         &&&&&&&&&&&&&& &&&&&&&&&&&&&\\
\hline\hline
\end{array}
\]
\caption{The lowest representations of the lower corner of the particle 
operator spectrum in the ${\sf X}/{\sf Y}$ plane (sector $\pi=+$). 
The states carrying quantum numbers of known elementary fermions organizes 
in a diamond in correspondence to the lightest particles of the spectrum.}
\end{table}
To large extracharge values correspond higher and higher --and presumably
heavier-- $su(2)$ and $su(3)$ representations. Simple isospin and color
multiplets appear only in the very bottom of the spectrum. In particular,
particle states carrying the correct hypercharge, isospin, color and chirality
to be identified with known elementary matter fields all carry extracharge 
${\sf X}\leq2$. In Table 2 we report the lowest weight representations of 
the lower corner of the particle spectrum in the ${\sf X}/{\sf Y}$ plane.
Particles carrying quantum numbers of known elementary fermions 
are indicated by the name of the corresponding first generation particles.
They organizes in a diamond at the very bottom of the spectrum.
Unknown elementary fermions are indicated by a question mark.
If we accept the idea that particles acquire mass in reason of the values 
of their extracharge, hypercharge and the complexity of their isospin and 
color representations, the  diamond correspond to the lightest 
states of the spectrum. \\
The content of the first six rows of Table 2 coincides with the sector
$\varsigma=+$ of Table 1 (plus an anti right-handed electron coming from 
the states
${\sf n_w}=0$, ${\sf n_s}=3$ that also have extracharge ${\sf X}=1$).
Table 2 allows us  to fully appreciate  the fundamental symmetries 
of the particle operator. The spectrum contains exactly a family and 
an anti-family of known elementary fermions. These are organized inside the
diamond in rows going from bottom-left to top-right as: 
{leptons} $\nu_{eR}-(\nu_{eL},e_L)-e_R$,
{anti-quarks} $d_R^*-(d_L, u_L)^*-u_R^*$,
{quarks} $u_R-(u_L, d_L)-d_R$ and 
{anti-leptons} $e_R^*-(e_L, \nu_{eL})^*-\nu_{eR}^*$. \\
Moving along bottom-left/top-right diagonals amounts to add a
weak quanta to the state, $({\sf n_w}, {\sf n_s})\rightarrow
({\sf n_w}+1, {\sf n_s})$ ({\it eg}.~$\nu_{eR}$ 
corresponds to ${\sf n_w}=0$, ${\sf n_s}=0$, 
$(\nu_{eL},e_L)$ to ${\sf n_w}=1$, ${\sf n_s}=0$ and  
$e_R$ to ${\sf n_w}=2$, ${\sf n_s}=0$;
$u_R$ corresponds to ${\sf n_w}=0$, ${\sf n_s}=2$,
$(u_L, d_L)$ to ${\sf n_w}=1$, ${\sf n_s}=2$ and
$d_R$ to ${\sf n_w}=2$, ${\sf n_s}=2$; etc.).
The relative infinitesimal generator is $({\sf X}+{\sf Y})/2$.
At a space-time level it produces a $U(1)$ rotation of all weak directions.
Therefore, it is natural to expect that the effective breakdown of weak 
symmetry couples states aligned on bottom-left/top-right diagonals. \\
Moving in orthogonal directions --from bottom-right to top-left-- amounts
instead to add a strong quanta to the state, $({\sf n_w}, {\sf n_s})
\rightarrow ({\sf n_w}, {\sf n_s}+1)$.
The corresponding infinitesimal generator $({\sf X}-{\sf Y})/2$ produces
a $U(1)$ rotation of all strong directions. If the color symmetry remains
a good one, there is no reason to expect a mixing between states belonging
to different bottom-left/top-right diagonals. Moreover, the exact 
conservations of the strong part of 
$C\!P_\bot=C\!P_{weak}\otimes C\!P_{strong}$ 
further divides the spectrum in two sub-sectors corresponding to 
$C\!P_{strong}=\pm1$. The lowest energy sector, $C\!P_{strong}=+1$,
contains the first and third bottom-left/top-right diagonals; these 
correspond exactly to a fundamental family of elementary fermions 
--repeated in many generations. Observe that the qualitative 
expectation for masses goes in the correct order of neutrino, electron,
quark up and quark down.
The other sector, $C\!P_{strong}=-1$, contains an anti-family.

\subsection*{Higher particle states}
New particles are presumably aligned along diagonals with a fixed value of 
${\sf n_s}$. The lightest new states correspond to ${\sf n_s}=0$,
$2$ and $4$. The simplest case is clearly the first,
corresponding to isospin multiplets, color singlets. We report 
the content in multiplets of the first six states belonging 
to the diagonal  ${\sf n_s}=0$ in the plane ${\sf I}/{\sf Y}$
\[
\begin{picture}(400,240)(-48,-40)
\put(340,150){...}
\put(340,100){...}
\put(340,-30){etc.}
\put(282,152){$\left(\begin{array}{r}       0_L\!\\[-1.5mm]
                                       \!\!-1_L\!\\[-1.5mm] 
                                       \!\!-2_L\!\\[-1.5mm] 
                                       \!\!-3_L\!\\[-1.5mm] 
                                       \!\!-4_L\!\\[-1.5mm]
                                       \!\!-5_L\!
                                       \end{array}\right)$}
\put(282,87){$\left(\begin{array}{r}\!\!-1_L\!\\[-1mm] 
                                    \!\!-2_L\!\\[-1mm] 
                                    \!\!-3_L\!\\[-1mm] 
                                    \!\!-4_L\!
                                    \end{array}\right)$}
\put(222,117){$\left(\begin{array}{r}     0_R\!\\[-1mm] 
                                     \!\!-1_R\!\\[-1mm] 
                                     \!\!-2_R\!\\[-1mm] 
                                     \!\!-3_R\!\\[-1mm] 
                                     \!\!-4_R\!
                                     \end{array}\right)$}
\put(222,58){$\left(\begin{array}{r}\!\!-1_R\!\\[-1mm] 
                                    \!\!-2_R\!\\[-1mm] 
                                    \!\!-3_R\!\\[-1mm]
                                    \end{array}\right)$}
\put(162,87){$\left(\begin{array}{r}\    0_L\!\\[-1mm] 
                                    \!\!-1_L\!\\[-1mm] 
                                    \!\!-2_L\!\\[-1mm] 
                                    \!\!-3_L\!
                                    \end{array}\right)$}
\put(162,28){$\left(\begin{array}{r}\!\!-1_L\!\\ 
                                    \!\!-2_L\!
                                    \end{array}\right)$}
\put(102,57){$\left(\begin{array}{r}    0_R\!\\[-1mm]
                                   \!\!-1_R\!\\[-1mm]
                                   \!\!-2_R\!
                                   \end{array}\right)$}
\put(117,0){${\mathbf e_R}$}
\put(43,28){$\left(\begin{array}{r}\!{\mathbf \nu\!\!\!\nu_{eL}}\!\!\\
                                      \!{\mathbf e_L}\!\end{array}\right)$}
\put(-5,0){${\mathbf\nu\!\!\!\nu_{eR}}$}
\put(285,-30){${\sf Y}={5\over2}$}
\put(225,-30){${\sf Y}=2$}
\put(165,-30){${\sf Y}={3\over2}$}
\put(107,-30){${\sf Y}=1$}
\put(49,-30){${\sf Y}={1\over2}$}
\put(-12,-30){${\sf Y}=0$}
\put(-50,150){${\sf I}={5\over2}$}
\put(-50,120){${\sf I}=2$}
\put(-50,90){${\sf I}={3\over2}$}
\put(-50,60){${\sf I}=1$}
\put(-50,30){${\sf I}={1\over2}$}
\put(-50,0){${\sf I}=0$}
\put(-53,-30){${\sf n_s}=0$}
\put(-57,-37){\line(1,0){420}}
\put(-57,-15){\line(1,0){420}}
\put(-18,-37){\line(0,1){230}}
\put(-57,-37){\line(0,1){230}}
\put(10,10){\line(5,3){30}}
\put(131,81){\line(5,3){30}}
\put(251,153){\line(5,3){30}}
\put(110,8){{\line(-3,1){30}}}
\put(164,45){{\line(-3,1){20}}}
\put(161,35){{\line(-3,1){20}}}
\put(225,80){{\line(-3,1){20}}}
\put(223,70){{\line(-3,1){18}}}
\put(222,59){{\line(-3,1){19}}}
\put(285,115){{\line(-3,1){20}}}
\put(282,105){{\line(-3,1){17}}}
\put(282,95){{\line(-3,1){17}}}
\put(281,83){{\line(-3,1){19}}}
\end{picture}
\]
Unknown particles are indicated by their electric charge after symmetry 
breaking. Solid lines indicate a possible pattern of symmetry breaking. 
Observe that the highest state with hypercharge ${\sf Y}=1$,
the lowest with ${\sf Y}=3/2$ as well as the highest with ${\sf Y}=2$ and 
the lowest with ${\sf Y}=5/2$ are in the sector $\varsigma=-$. If $\varsigma$
is a relevant selection rule the first new observable fermions are presumably
a mixing of the higher ${\sf Y}=3/2$ with the lower ${\sf Y}=2$ states.

\section{Particle-Energy relation, extraforce and  ${\mathbf \nu\!\!\!\nu_R}$} 

All what has been discussed in this paper is a direct consequence of a 
single basic assumption:
local gauge invariance associated to {\it all} fundamental interactions 
reflects observer's freedom of choosing a space-time reference frame 
in every point. \\
 Given this, the gauge group associated to gravitational, electromagnetic, 
weak and strong interactions determines the local geometry of space-time 
--including its dimension. 
The geometrical framework naturally embodies degrees of freedom 
corresponding to gravitational and non-gravitational gauge interactions.
 In order to make a first connection between theory and observation,
in this paper we considered the free motion of a test spinor in a given 
background. As a consequence of the local geometrical structure the fermion 
is reduced on a $1+3$ effective space-time; the effective four dimensional 
dynamics is described by chiral fermions carrying quantum numbers of known 
elementary particles --including generations; matter couples to gauge forces 
in the standard way.
The fermionic sector of the Standard Model of Elementary Particle 
emerges as the low energy limit of the theory. \\ 
 Of course, before claiming that we have a unified comprehension of
fundamental interactions and elementary matter fields, there are two
major issues that have to be investigated:
the fundamental equations relating space-time geometry and matter 
distribution --the extension of Einstein equations-- 
{\it and} 
the analysis of a natural mechanism producing electro-weak symmetry breaking.
In the low energy limit, this should bring us to an understanding 
of the splitting of coupling constants associated to fundamental interactions
as well as to a first principle prediction of masses and mixing angles.
Both subjects are currently under investigation. For now, we observe that
already at the stage considered in this paper the theory leads to  
important phenomenological consequences (other than the predictions of 
new particles and generations):

\subsection*{Particle-Energy relation}
In the very fast rotation of 
the fourteen dimensional spinor around ordinary space-time is stored 
an amazing amount of energy. This appears as a fourteen dimensional 
mass term in matter field equations. Therefore, in close analogy to 
the special relativistic picture associating to every body of mass $m$ 
an intrinsic energy $E_m=mc^2$, our theory predicts an even more fundamental 
relation: to every elementary particle $\psi_{\sf p}$ is associated an 
intrinsic energy
\[
E_{\sf p}={\hbar c\over l}{\sf p}
\]
A fundamental length of the order of the Planck scale $l\approx10^{-33}\ cm$
associates to the electron an intrinsic energy of about $10^{15}\ TeV$.
This enormous amount of energy stored in every elementary particle does
not affect directly the effective four dimensional 
dynamics\footnote{In many respects the situation is analogous to that 
of many adiabatic systems. To fix ideas we may think of the separation of 
vibrational and rotational nuclear freedoms in a molecule 
(with no symmetries). Most of the energy  is stored in a vibrational state. 
In correspondence the molecule performs a free rotation. On rotational energy 
scales there is no direct perception of the amount of energy stored in 
vibrations. In addition, vibrational transitions are too rare to be 
appreciated on rotational energy scales.}.
 However, it affects it indirectly. 
The ability of a particle of carrying extracharge, hypercharge, isospin and 
color is proportional to its intrinsic energy.
 A more striking consequence is that every elementary particle but $\nu_R$ 
is unstable. In the effective four 
dimensional theory, the conservation 
of extracharge, hypercharge, isospin and color prevent the system from
collapsing down into the ground state. However, on very long time scales
--or very high energies-- we should expect these transitions to become 
relevant.

\subsection*{Extraforce}
Enforcing the gauge group of fundamental interactions at a space-time 
level indicates the existence of a further $U(1)$ gauge symmetry. 
This is presumably associated to an additional force mediated by an
additional gauge boson. Predictions on the behavior of the extraforce 
and on the mass of the corresponding vector boson go necessarily through the 
extension of Einstein equation to the whole fourteen dimensional space-time 
and through the analysis of symmetry breaking.

\subsection*{$\nu_R$}
The theory predicts the existence of a particle carrying the right quantum 
numbers to be identified with the right-handed neutrino $\nu_R$.  
It is the ground state of the square of the particle operator ${\cal P}$; 
the only elementary particle carrying a vanishing intrinsic energy. 
In some sense, therefore, it represents the most elementary particle 
we can think of in four dimensions. 
The right-handed neutrino carries zero extracharge, zero hypercharge, 
no isospin and no color. It has no electromagnetic, weak, strong nor
extra interactions.
It only interacts through the gravitational field. It is natural to expect 
that after electro-weak symmetry breaking  $\nu_R$ couples to $\nu_L$ 
originating the lightest massive particle. In addition it is natural to
expect that almost the totality of the matter in the universe is in the 
ground state $\nu_R$. This could possibly give an answer to the dark matter 
problem. 

\newpage
\section*{\hspace{4mm} Considerations and Acknowledgments}
 The idea that the fermionic sector of the Standard Model could emerge
as a low energy limit from an extension of space-time relativity came first
to me when I realized that the effective motion of a generic dynamical 
system constrained on a sub-space by a scalar potential gets naturally  
coupled to an $SO(n_1)\times SO(n_2)\times ...$ gauge potential. 
Dimension and number of $SO(n)$ factors depend on subspace 
codimension and symmetries of the confining  potential (a beautiful example 
is found in the separation of vibrational and rotational 
freedoms in polyatomic molecules). A gauge group given 
by the direct product of orthogonal groups is not that different from 
a gauge group given by the direct product of unitary groups.
The possibility that space-time dimensional reduction is produced by a 
scalar potential is unappealing. However, the property of magnetic fields of
effectively reducing the motion along field lines seemed to match
perfectly the unitary structure of non-gravitational gauge groups.
 I started working on the idea of a mixed real-complex geometry
when I was  `Bruno Rossi' Post Doc fellow at MIT.
 There I first presented the idea of ``Dimensional Reduction without 
Compactification'' in a seminar in April 1998. 
 I kept on working on the problem at  KFKI-RMKI in Budapest 
--I take the chance to thank Prof.~K.~T\'oth for that position-- where 
I first obtained the fermionic spectrum summarized in Table 1. 
The work was reorganized and completed at the Physics Department
of the University of Parma where I spent the winter 2000 as an 
external collaborator. \\
It is a genuine pleasure to thank my friends Roberto De Pietri,
L\'aszl\'o Szabados and P\'eter Vecserny\'es 
for all the help they gave me. Discussing with 
them the many aspects of the problem was precious to me.


\end{document}